\documentclass[12pt]{JHEP3}  
\usepackage{amsfonts}    
\usepackage{amsmath}    

\def\a{\alpha}  
\def\b{\beta}  
\def\g{\gamma}  
\def\l{\lambda}  
\def\d{\delta}  
\def\e{\epsilon}  
\def\t{\theta}

\def\G{\Gamma}

\def\p{\partial}

\newcommand{\be}{\begin{equation}} 
\newcommand{\ee}{\end{equation}} 
\newcommand{\bea}{\begin{eqnarray}}  
\newcommand{\eea}{\end{eqnarray}} 
\newcommand{\nn}{\nonumber}

\title{\Large Covariant One-Loop Amplitudes in D=11}    
\author{Lilia Anguelova  
\\ Michigan Center for Theoretical Physics, University of Michigan\\  
Ann Arbor, MI 48109-1120, USA, \\ 
anguelov@umich.edu}    
\author{Pietro Antonio Grassi\\ C.N. Yang Institute for Theoretical Physics,   
SUNY at Stony Brook\\  
Stony Brook, NY 11794-3840, USA, \\and \\ 
DISTA, Universit\`a del Piemonte Orientale, \\
Piazza Amborsoli, 1 15100 Alessandria, Italy\\
pgrassi@insti.physics.sunysb.edu}    
 \author{Pierre Vanhove\\ 
CEA/DSM/SPhT, URA au CNRS, CEA/Saclay,\\
 F-91191 Gif-sur-Yvette, France\\ 
 vanhove@spht.saclay.cea.fr}   

\abstract{   
We generalize to the eleven-dimensional superparticle Berkovits' prescription for loop computations in the pure spinor approach to covariant quantization of the superstring.
Using these ten- and eleven-dimensional results, we compute covariantly the following
one-loop amplitudes:  $C\wedge X_8$ in M-theory;  $B\wedge X_8$ in type II string theory and ${\cal F}^4$ in type I.
We also verify the consistency of the formalism in eleven dimensions by
recovering the correct classical action from tree-level amplitudes.
As the superparticle is only a first approximation to the supermembrane, we comment on the possibility of 
extending this construction to the latter.
Finally, we elaborate on the relationship between the present BRST language and the spinorial cohomology approach to corrections of the effective action.
}  

\preprint{ 
MCTP-04-46,    
YITP-04-41\\
SPHT-T-04/102,
NEIP-04-04\\
hep-th/0408171 
}

\keywords{pure spinors, superparticle, eleven dimensions} 
\begin{document} 

\section{Introduction} 
Since the discovery of the web of dualities that relate the five 10-dimensional
string theories with each other and with 11-dimensional supergravity
\cite{Mtheory}, it has been realized that perturbative and nonperturbative
effects in ten-dimensions are encoded in the effective action of an underlying
11-dimensional theory. In particular, both the fundamental string and the
D$2$-brane originate from the 11-dimensional supermembrane. Although the
relationship between this latter M$2$-brane and the eleven-dimensonal
supergravity fields is what one would expect from a fundamental object and its
long-range fields \cite{BST}, it has turned out that there is no gap between the
massless and massive exitations of the membrane
\cite{deWit:1988ct,Dasgupta:2002iy}. Nevertheless one can consistently define
vertex operators for the massless fields \cite{Dasgupta:2000df} and study the
corresponding scattering amplitudes.\footnote{On the other hand, vertex
operators, analogous to the massive superstring states, are not expected to
exist due to the continuous spectrum of the supermembrane. Instead, the membrane
excitations are interpreted as being related to multi-particle (rather than
one-particle) states in the Matrix Theory proposal of \cite{BFSS}.} In full
generality, this is still a task beyond reach. However in the limiting case of
the superparticle, i.e. suppressing the transverse fluctuations of the
supermembrane, several successful computations have been performed in the light
cone gauge. More precisely, the 11d four-graviton scattering was computed to one
and two loops in \cite{GGV, GGK} and \cite{GKP} respectively. Of course, these
supergravity amplitudes are UV divergent but the lack of knowledge about the
microscopic degrees of freedom can be compensated by information provided from
various dualities. This is how, for example, the one-loop generated
eleven-dimensional $R^4$ term was ``renormalized" in \cite{GGV}. The important
lesson from the point-particle limit of the membrane is that quantum
supergravity in 11d does indeed reproduce the correct form of certain terms in
the effective action of M-theory, for example the 11d one-loop $R^4$ term which
gives rise to both the tree level and the one-loop $R^4$ terms in the effective
superstring action.

However, the light cone gauge, used in the above works, has its drawbacks in
that it is not suitable for computing all possible amplitudes in a generic
background and also it is less efficient in organizing the calculations, in
particular making manifest the cancellations between different contributions,
than a covariant formulation would be. The first steps towards a full covariant
description were made in \cite{membr}, based on the recent progress in the
covariant formulation of 10d superstring theory \cite{Berkovits:2000fe}. The
idea is to construct a supermembrane action, alternative to \cite{BST}, which
reduces to the ten-dimensional pure spinor description of the type IIA
superstring after double dimensional reduction. This action is invariant under a
BRST transformation generated by the same type of operator $Q=\l^A d_A$ as the
one used in \cite{Berkovits:2000fe} for the case of superstrings. So again one
can define vertex operators via the cohomology of $Q$. Using this covariant BRST
formulation, the general form of the vertex operators relevant for the
supermembrane theory massless excitations, namely the graviton $g_{MN}$,
gravitino $\psi^A_M$ and three-form $C_{MNP}$, was obtained in \cite{membr},
although some of the superfields in that expression remained undetermined.
Clearly, these vertices can be reduced to those of the 11d superparticle by
setting to zero the transverse fluctuations of the supermembrane.

Although \cite{membr} gives a tree-level prescription for integration over the
zero modes of the pure spinor ghosts $\l^A$ and their conjugate momenta $w_A$ it
does not derive in general the measure for this integration and so leaves
performing loop computations out of reach. The same problem was resolved only
very recently for the pure spinor formulation of 10-dimensional string theory
\cite{Berkovits:2004px}. The crucial ingredient was the construction of a
composite field playing a role analogous to the one of the NSR $b$-antighost.
This is very nontrivial in Berkovits' formalism as the only field with negative
ghost number, $w_A$, appears always in combination with $\l^A$, i.e. only in
expressions with ghost number zero. This problem was overcome in
\cite{Berkovits:2004px} by introducing picture-changing operators and defining
the $b$-ghost via $\{Q, b \} = T Z$, where $T$ is the energy-momentum tensor and
$Z$ is a picture raising operator with ghost-number $1$. As pointed out in that
work, the necessity of picture-changing operators should have been anticipated,
since the pure spinor ghosts are world-sheet bosons with zero modes similarly to
the NSR $(\beta, \gamma)$-ghosts.

Unlike the superstring, the membrane world-volume theory is not conformally
invariant. But, as in the string case, it is still reparametrization invariant.
Therefore one should expect an analogue of the $b$-ghost in eleven dimensions as
well. In this paper we show that the 10-dimensional covariant rules for
superstring loop calculations of \cite{Berkovits:2004px} can be extended to the
case of the 11-dimensional superparticle. In fact, we construct the 11d
generalization of the ``antighost'' field $b_{B}$ in a form that is also valid
for string theory. The advantage of our different, though equivalent to the one
in \cite{Berkovits:2004px}, representation of $b_B$ is that it simplifies the
prescription for one loop computations given there. Our construction of
covariant one loop amplitudes, as well as the light-cone calculations of
\cite{GGK}, is based on the string-inspired first quantized approach
to field theory developed in \cite{Strassler:1992zr, BK,GGV, GKP}. Namely, one inserts
vertex operators, describing the external fields, on the world-line of the
particle circulating in the loop and integrates over that world-line. For that
purpose we derive the 11d superparticle vertex operator and show that it is
consistent with the general form obtained in \cite{membr}.\footnote{Recall that,
as we already mentioned, many of the superfields in that general form are
actually still unknown.}

Using the superstring formalism of \cite{Berkovits:2004px} and the 11d one of
this paper, we compute covariantly several one-loop amplitudes: \,the 
terms $B\wedge X_8$ in IIA/B string theory and $C_{3}\wedge X_8$ in 11d
supergravity
and also the supersymmetric ${\cal F}^4$ in type I (meaning all three cases: four
gluons; two gluons and two gluinos; four gluinos), always finding the correct
answers. We also show that the tree-level scattering amplitudes in 11d reproduce
completely the two-derivative action of eleven-dimensional supergravity
\cite{CJS}. We view our results as a strong confirmation of the viability of the
pure spinor approach to covariant quantization.

Finally, we elaborate on the relashionship/equivalence between pure spinor
cohomology and the spinorial cohomology developed in
\cite{swedes,CNT,CNT11d,HoweTsimpis}, recovering the latter from an extended
BRST operator similarly to the construction of \cite{Grassi:2001ug} that relaxes
the pure spinor constraint. 
 
We conclude with some remarks about a possible extension of the present computation to the M$2$-brane.



\section{From $\kappa$-symmetry to BRST symmetry for 11d superparticle}
\label{section:map} 

To introduce basic notions, we review in the current section the path leading from the $\kappa$-symmetric superparticle action of \cite{Brink:1981nb} to the action of Berkovits \cite{Berkovits:2001rb}, essentially following \cite{ICTP}. We also explain how the tree-level measure in the pure spinor formalism, introduced initially ad hoc, maps to the light-cone measure.
 
The Brink-Schwarz super-Poincar\'e invariant point-particle action is
\begin{equation} 
S_{covariant} = \int d\tau \, \left( P_M \Pi^M -\frac12 e P_M P^M\right) \, ,
\end{equation} 
$$ 
\Pi^M = \dot X^M- i\bar\Theta\Gamma^M \dot\Theta \, ,
$$ 
where $M=0,\dots,10$, \,$\tau$ parametrizes the superparticle worldline, $\Theta^{A}$ are Majorana spinors with 
$A=1, \dots, 32$, \,$e$ denotes the worldline einbein and the 
indices of $P^{M}$ are lowered by the flat metric. 
This action is invariant under global supersymmetry, 
\be 
\delta_\epsilon \Theta = \epsilon; \qquad \delta_\epsilon X^M = i\bar\epsilon\Gamma^M\Theta; \qquad \delta_\epsilon P_M=\delta_\epsilon e=0 \, ,
\ee 
and local $\kappa$-symmetry,
\be
\delta_\kappa \Theta = iP_M\Gamma^M \kappa; \qquad \delta_\kappa X^M = i\bar\Theta\Gamma^M \delta_\kappa \Theta; \qquad \delta_\kappa e = 4e \dot{\bar\Theta}\kappa; \qquad \delta_\kappa P_M=0 \,.
\ee 
It is also invariant under local reparametrizations. 

Using $\kappa$-symmetry one can eliminate half of the components of $\Theta$. This is easily seen by choosing the light-cone gauge 
\cite{GGK} 
$$ 
X^+ = x^+ + p^+ \tau;\qquad \Gamma^+\Theta=0 \, , 
$$ 
where 
$$ 
X^\pm = \frac{1}{\sqrt2} \left(X^0\pm X^{10} \right);\qquad \Gamma^\pm = \frac{1}{\sqrt2} \left( \Gamma^0 \pm \Gamma^{10}\right),\quad \{\Gamma^+, \Gamma^- \}=1 \, .
$$ 
The light-cone gauge action
used by Brink and Schwarz \cite{Brink:1981nb} is
\begin{equation}\label{lc} 
S_{l.c.} = \int d\tau\, \left( \dot X^M P_M + \frac{1}{2} \dot S^\alpha S_\alpha- 
\frac12 e P^M  P_M \right) ,
\end{equation} 
where $S^\alpha=\sqrt{p^+}(\Gamma^- \Theta)^\alpha$ are the sixteen surviving SO(9) components of $\Theta$. In this gauge half of supersymmetry is realized linearly and the other half - nonlinearly: 
\begin{eqnarray}\label{linear} 
\delta_\varepsilon S^{\a} &=& \sqrt{p^+} \, \varepsilon^{\a}; \qquad \delta_\epsilon X^M= 0;  \qquad\delta_\varepsilon P_M=\delta_\varepsilon e=0 
\\ 
\label{nonlinear}\delta_\eta S &=& \frac{i}{\sqrt{p^+}} P_M\Gamma^M\Gamma^+\eta; \qquad \delta_\eta X^M =-\frac{2}{\sqrt{p^+}} \bar\eta \Gamma^M\Gamma^+ S; \qquad \delta_\varepsilon P_M=\delta_\varepsilon e=0 \, ,
\end{eqnarray} 
where the parameter of the former, $\varepsilon$, satisfies $\Gamma^+\varepsilon=0$ and the parameter of the latter
--- $\Gamma^+ \eta\neq 0$. Note also that in order to preserve $\Gamma^+\Theta=0$
one has to combine a linear supersymmetry transformation with a $\kappa$-symmetry one.

The wavefunction of the system carries SO(9) vector and fermionic degrees of freedom and can be expanded on the basis of physical states: the metric  $|IJ\rangle$, 3-form $|IJK\rangle$ and gravitino $|\a I\rangle$ (see \cite{GGK} for details), where $I,J,K = 1,...,9$. So the superspace is reduced from 32 original Grassmann 
coordinates to 16 free ones, $S_{\a}$, 
satisfying the Clifford relations $\{S_{\alpha}, S_{\beta}\} = 
\delta_{\alpha \beta}$. 
At tree level in string expansion,  
they can be organized into $SO(8)$ spinor representations 
$S_{\a} = (\xi_{a}, \xi^{\dagger}_{a})$ and the tree level measure can be defined by the path integral formula
\begin{eqnarray}\label{so8}
\langle  \xi_{a_{1}} \dots \xi_{a_{8}} \rangle_{g=0} = 
\int d^{8}\xi \prod_{i=1}^{8} \xi_{a_{1}} \dots \xi_{a_{8}} 
= \e_{a_{1} \dots a_{8}}\,.
\end {eqnarray}
Accordingly, the states $|IJ\rangle$, $|IJK\rangle$ and $|\a I\rangle$ 
are given in terms of $SO(8)$ representations obtained by acting with $\xi_{a}$ on the vacuum defined by $\xi^{\dagger}_{a} |0\rangle =0$ with 
$a\neq 0$. This agrees with the fact that at 
tree level only 8 zero modes coming from $\xi_{a}$ 
have to be saturated ($\xi^{\dagger}_{a}$
play the role of conjugate momenta, i.e. $\{\xi_{a}, \xi^{\dagger}_{b}\} = \delta_{ab}$). Alternatively, we can write the one loop measure in a completely $SO(9)$ covariant way by saturating both   
the zero modes of $\xi_{a}$ and those of 
$\xi^{\dagger}_{a}$:
\begin{equation}\label{SixteenTheta} 
\left\langle S^{\alpha_1}\cdots S^{\alpha_{16}}\right\rangle_{g=1} =
\int d^{16}S \prod_{i=1}^{8} S^{a_{1}} \cdots S^{a_{16}} 
=\epsilon^{\alpha_1\cdots \alpha_{16}} \,.
\end{equation} 

In order to compare (\ref{lc}) with the pure spinor approach, one 
adds two new doublets $(p_A,\theta^A)$ and $(w_{A}, \lambda^{A})$, of 
anticommuting and commuting variables respectively, which are 
not related to the light-cone fermions $S^\alpha$. Then one defines 
\begin{equation}\label{dhat} 
\hat d_A = p_A + \frac{1}{2} P_M \left[\Gamma^M\left(\theta+\frac{1}{\sqrt{p^+}} \Gamma^+ S\right)\right]_A \, ,
\end{equation} 
which satisfies  
$$ 
\{ \hat d_A , \hat d_B\}= -\frac{i}{2p^+} P_M P^M (\Gamma^+)_{AB}\ . 
$$ 
In deriving this relation we have used that $S^\a$ are their own canonical conjugates (i.e., they satisfy $\{S_\alpha, S_\beta\}= \delta_{\alpha\beta}$). 
The first-class constraints $d_A \approx 0$ {}\footnote{Notice that 
$P_{M}P^{M} \approx 0$ is also a first class constraint.} 
are associated with the BRST operator 
\begin{equation}\label{Qprime} 
Q'  = \lambda^{\prime A} \hat d_A; \qquad \lambda'\Gamma^+ \lambda' = 0\ , 
\end{equation} 
and the gauge-fixed action 
\begin{eqnarray}\label{ActionI} 
S_{cov} = \int d\tau \Big( P_{M} \dot X^{M} - \frac12 e P_M P^M + \dot \theta^A p_A + \frac{1}{2}\dot S^\alpha  S_\alpha  + \dot \lambda^{\prime A} w'_A \Big)\,. 
\end{eqnarray} 
In order to bring the BRST charge (\ref{Qprime}) into the form of Berkovits' BRST charge  
\begin{equation}\label{Qb} 
Q = \lambda^A d_A \qquad \text{with}\qquad \lambda\Gamma^M \lambda=0 \, ,
\end{equation} 
we split the SO(10,1) commutting spinor $\lambda^{\prime}$ into its SO(9) 
components with respect to $\Gamma^+$ as (see Appendix~\ref{SOeight} for more details) 
$$ 
\lambda^{\prime A} = (\lambda'{}_1^\alpha,\lambda'{}_2^\alpha) \, .
$$ 
With this decomposition the BRST charge (\ref{Qprime}) becomes 
$$ 
Q'= \lambda'{}_2^\alpha d_{2\alpha} + \lambda'{}_1^\alpha \hat d_{1\alpha} \, , 
$$ 
where we have used the subspace of SO(9) spanned by $\Gamma^+ \lambda' 
=0$, $\hat d_{2\alpha}  = d_{2\alpha}$. Since 
$\lambda'{}_1^\alpha=(\Gamma^+\lambda')^\alpha=((\Gamma^+\lambda')^a, (\Gamma^+\lambda')^{\dot a})$ 
is a null SO(9) spinor preserved by a SO(8) subgroup chosen to act on the indices 
$\alpha$, there is a similarity transformation \cite{ICTP} that sends the components 
$\left(\Gamma^+S\right)_a$ onto $\left(\Gamma^+d\right)_a$: 
\begin{equation}\label{simil} 
Q' \to e^{-i  S\Gamma^+d /\sqrt{p^+}}\ Q'\ e^{i  S\Gamma^+d/\sqrt{p^+}} 
\end{equation} 
  and the components 
$\left(\Gamma^+ S\right)_{\dot a}$ are re-absorbed by shifting 
$\left(\theta+\sqrt{p^+}^{-1}\, \Gamma^+S\right)_{\dot a}\to \theta_{\dot a}$ in 
(\ref{dhat}). The similarity transformation has to preserve the null spinor 
  nature of $\Gamma^+\lambda'$ and one component of   $(\Gamma^+S)^a$ is not 
  gauged into $d_a$ by (\ref{simil}) but is shifted into an extra component of $\theta$. As shown in Appendix~\ref{SOeight}  the components of $\lambda$ are identified with the components of $\lambda'$ via 
  $\lambda^A=(\lambda^{\prime\dot a},\sqrt{v^+-s_is^i},s_i, 0)$ with 
  $i=1,\cdots, 7$, which gives $16+8-1=23$ components. 

The final result is the BRST invariant action 
\be\label{BRSTaction}
S_{cov} = 
\int d\tau \Big( P_{M} \dot X^{M} - \frac12 P_M P^M + \dot \theta^A p_A + 
\dot \lambda^{A} w_A \Big)\,. 
\ee
where $\l^{A}$ satisfies the pure spinor constraint 
$\l^{A} \left(\G^{M}\right)_{AB} \l^{B} =0$.\footnote{Note that this definition differs form Cartan's pure spinors, also used by P. Howe, since the latter obey in addition $\l^{A} \left(\G^{MN}\right)_{AB} \l^{B}=0$. Interestingly enough, similar constraints, more precisely $\l^{A} \left(\G^{MN}\right)_{AB}\l^{B}\, \Pi_{M}=0$, emerge in the quantization of the supermembrane \cite{membr} and the covariant description of D-branes \cite{lilia-antonio}. Another obvious remark about the 11d constraint $\l^{A} \left(\G^{M}\right)_{AB} \l^{B} =0$ is that when written in terms of two 10d pure spinors it coincides exactly with the two conditions for the D$0$-brane \cite{lilia-antonio}, which should be the case since the latter is related to the 11d superparticle by dimensional reduction.} 
The BRST operator becomes $Q= \l^{A} d_{A}$, which is 
nilpotent only up to pure spinor constraints and gauge transformations 
$\delta w_{A} = \Lambda_{M} (\Gamma^{M} \l)_{A}$, where 
$\Lambda_{M}$ is a local gauge parameter. It is easy to check that 
$S_{cov}$ is invariant under $\delta w_{A}$ and that also $\dot Q = 0$. Notice that on-shell, the only non constant field is $X^{M}$. The action $S_{cov}$ is manifestly super-Poincar\'e invariant. In addition, the ghost field part of the Lorentz generators, $N^{MN} = \frac12 w \G^{MN} \l$, as well as the ghost current $J= w_{A} \l^{A}$ are 
invariant under the gauge transformations of $w_{A}$. 

The light-cone variables satisfy the zero mode
saturation identity (\ref{so8}), which expresses the fact that the superparticle preserves half of the original supersymmetry. We will see now that it gets mapped into the tree level measure
\begin{equation}\label{normtree}  
\langle (\lambda\Gamma^{M_1}\theta)\cdots (\lambda\Gamma^{M_7}\theta)\,  (\theta\Gamma_{M_1\cdots M_7}\theta) \rangle_{g=0} = 1 
\end{equation} 
in the pure 
spinor formalism. For that purpose, we work in the space 
with both species of variables and 
we define the vertex operator $U^{(3)} = \l^{A} \l^{B} \l^{C} A_{ABC}(\t)$ 
by its expansion in terms of the background fields $C_{MNP}, g_{MN}$ and $\psi_{AM}$. Now, we use the expressions for $C_{MNP}, g_{MN}$ and 
$\psi_{AM}$ in terms of the light-cone $\xi_{a}$-expansion and so obtain a vertex operator of the 
form $U^{(3)} = \l^{A} \l^{B} \l^{C} A_{ABC}(\t, \xi)$. In the 
same way we have to define a vertex operator $U^{(4)}$ (see 
\cite{membr} and next sections for details). Then one can show that 
\begin{equation}\label{conv}
\e^{a_{1} \dots a_{8}} \xi_{a_{1}} \dots \xi_{a_{8}} =  \langle 
U^{(3)}(\l,\t, \xi) U^{(4)}(\l,\t,\xi) \rangle_{g=0} \,\, ,
\end{equation}
where the correlator is taken only in the $\l, \t$ space. This 
maps the measure for zero modes in the pure spinor approach 
into the measure factor for light-cone variables at tree level.

 Concluding this section, we note that from the above light-cone discussion it follows that any
 correlator with more than sixteen fermions vanishes. This means that any tree-level
 correlators $\langle \lambda^{7+p} \theta^{9+q}\rangle$, with $p$ and $q$
 positive integers, are zero. In other words, the zero momentum cohomology of $Q$ (see (\ref{Qb})) is empty for $\lambda$-ghost number bigger than 7, as shown by a direct computation in \cite{membr}. The same reasonning can be applied for the ten dimensional open string to infer that the zero momentum cohomology of the corresponding $Q$ is empty for $\lambda$-ghost number greater than 3.

\section{Supergravity in 11 dimensions and vertex operators} 

In this section we analyze the structure of the vertex operators for the particle limit of the eleven-dimensional supermembrane. First we derive them from BRST cohomology and explain their component field expansion. We also point out that they are consistent with the general expression with undetermined coefficients of the integrated covariant vertex of \cite{membr}. Then by coupling to a nontrivial background we derive another version of the integrated and unintegrated operators, which will be useful in the rest of the paper.

From now on we use the following conventions in 11d: $M,N,...$ $-$ curved vector indices; $A,B,...$ $-$ curved spinor indices; $r,s,...$ $-$ flat vector indices; $a,b,...$ $-$ flat spinor indices. Also $Z^{\Omega} \equiv (X^M, \Theta^A)$. In 10d we denote by $m,n,...$ curved vector indices and by $\a, \b,...$ curved spinor indices. Finally $I,J,...=1,2$ run over the spatial directions of the membrane and $\partial_0$ is a derivative w.r.t. the M2 worldvolume time.


\subsection{Vertex Operators and $N=1, D=11$ Supergravity}\label{section:vop} 

The vertex operator for the 11d superparticle $V^{(0)}=\int d\tau \, \mathcal{V}^{(0)}$ at ghost number zero can be viewed as a deformation of the quantum action expanded around flat background. It emerges as the first non-trivial contribution 
\begin{equation}\label{defactioA}
S_{def} = S_{cov} + \int  d\tau \, \mathcal{V}^{(0)} + {\cal O}(V)\,,
\end{equation} 
where ${\cal O}(V)$ denotes higher terms in the expansion. In order that the action $S_{def}$ is still invariant under BRST symmetry, one has 
$\{Q, \mathcal{V}^{(0)} \}  = \partial_{\tau} U^{(1)}$, where $ U^{(1)}$ is a ghost number one vertex operator. The latter satisfies $\{Q, U^{(1)}\} =0$ by consistency. 

Acting with the BRST operator $Q$ on a generic polynomial of the fields $P_{M},$ $X^{M},$ $d_{A},$ $\t^{A},$ $\l^{A}$ and $N_{NR} = \frac{1}{2}(w\Gamma_{NR}\lambda)$, one finds that
\bea \label{VO}
\mathcal{V}^{(0)} = P^{M} \Big( g_{MN} P^{N} + E_M{}^A d_A +  
H_{M}^{[NR]} N^{NR}  \Big) 
\eea 
and
\be \label{vopI}
U^{(1)} =\lambda^{A}E_{A}{}^{r} E_{M}{}^{s} P^{M}\, \delta_{rs} \, ,
\ee
where $g_{MN} = E_{M}{}^{r} E_{N}{}^{s} \delta_{rs}$\,, \,$E_{M}{}^{A}$ and $H_{M}^{[NR]}$ are superfields which satisfy the supergravity field equations. We also note that (\ref{VO}) is easily seen to be consistent with the light-cone vertices of \cite{GGK}.
  
The vertex operator $\mathcal{V}^{(0)}$ is defined up to a gauge transformation generated by total derivative terms, i.e. $\delta \mathcal{V}^{(0)} = \partial_{\tau} \Omega^{(0)}$. Correspondently, the vertex $U^{(1)}$ is defined up to BRST exact terms of the form $\delta U^{(1)} = \{Q, \Omega\}$. For the case at hand, the gauge transformations are given by
\bea\label{gg} 
\Omega = \Omega^{M} P_{M} +  
\Omega^{A} d_{A} + \Omega^{[MN]} N_{MN}\, .
\eea 
At leading order in the $\theta$ expansion their parameters are the following:  
\bea\label{ccgg} 
\Omega^{M} = \xi^{M} + {\cal O}(\t) \,, ~~~~~ 
 \Omega^{A} = \epsilon^{A} + {\cal O}(\t) \,, ~~~~~ 
  \Omega^{[MN]} = \omega^{[MN]} + {\cal O}(\t) \, ,   
\eea 
where $\xi^{M}$ is the one of infinitesimal diffeomorphisms,  
$\epsilon^{A}$ $-$ of infinitesimal supersymmetry transformations and  
$\omega^{[MN]}$ $-$ of Lorentz gauge transformations (i.e., $SO(10,1)$ rotations).  

The superfields $E=(E_{M}{}^{r}, E_{M}{}^{a},E_{A}{}^{r})$ and $H^{[NR]}_{M}$, contained in the vertex operator ${\cal V}^{(0)}$, are comprised from the fields of the usual 11d supergravity multiplet namely the graviton $e_{M}^{r}$, gravitino $\psi_{M}^{A}$ and 3-form $C_{MNR}$. However, the 11d superparticle does not couple directly to the potential $C_{MNP}$, but only feels its field-strength $G=dC$. As a result, the tensor gauge transformations cannot be seen among the transformation rules given in (\ref{gg}).  

The expansion of the superfields  
$E$, $H_{M[NR]}$  
up to order ${\cal O}(\theta^3)$  
in terms of the on-shell supergravity fields is the following \cite{KasperMembrane,DimitriosNew}: 
\begin{eqnarray} \label{superVielbein}
E_M{}^r &=& e_M{}^r +   
2 \bar\theta \Gamma^r \psi_M + \left(\bar \theta \mathcal{U}^{r| NPQL}{}_M\theta\right) G_{NPQL} + 
\frac{1}{4} \bar\theta \Gamma^{r\, st}\theta \omega_{M[st]}+{\cal O}(\theta^{3}) \nn\\ 
E_M{}^a &=&\psi_M{}^a + \frac{1}{4} \omega_{M\, [rs]} (\Gamma^{rs}\theta)^a + \left(\mathcal{T}_M{}^{NRST}\theta\right)^a
G_{NRST} + {\cal O}(\theta^{3}) \nn\\ 
E_A{}^r &=& -\left(\bar\theta \Gamma^r\right)_A + {\cal O}(\theta^{3}) \nn\\ 
H_{M}^{[rs]} &=& \omega_{M}^{[rs]} +  {\cal O}(\theta^{2})  
\end{eqnarray} 
with
\begin{eqnarray}\label{tensorCJST}
\mathcal{T}_M{}^{NRST}& = &\frac{1}{288}  
\left(\Gamma_M{}^{NRST} - 8 \eta_M{}^{[N} \Gamma^{RST]} \right); \\
 \label{tensorCJSU}\mathcal{U}^{R| 
 MNPQL}&= &\frac{1}{288} \left(5\eta^{R[M}\Gamma^{NPQL]}+8\Gamma^{R[MNP} \eta^{Q]L} \right) \,.
\end{eqnarray}
We note in passing that there are several different ways of obtaining the above expansions. The most well-known one is the  gauge completion method used in \cite{KasperMembrane}. A much easier alternative is the following: one can use the field equations and Bianchi identities
to derive recursion relations for the component fields in the $\theta$ expansion of the supergravity superfields 
similarly to the case of 10-dimensional SYM \cite{HS}. This procedure was applied recently to ten- and eleven- dimensional
supergravity in \cite{Grassi:2004ih} and \cite{DimitriosNew} respectively.\footnote{We should mention yet another way of
calculating the component fields in the $\theta$ expansion
of superfields, namely the normal coordinates approach advocated for example in \cite{GK}. }

The superspace description of 11d supergravity, obtained by starting from the ghost number zero vertex operator $\mathcal{V}^{(0)}$, coincides with the superspace description 
given in \cite{CF,BH}. However, as shown in \cite{Nishino:2001mb,swedes}, one can also use a formulation based on the constraint of the 
spinorial component of the four form $G_{4} = dC_{3}$. This constraint can be 
translated into a BRST closedness condition by considering another type 
of vertex operator $U^{(3)} = \l^{A} \l^{B} \l^{C} A_{ABC}$. 
The superfield $A_{ABC}$ is defined 
up to gauge transformations $\delta A_{ABC} = \G^{M}_{(AB} \Xi_{C)M}$.  By requiring 
that $U^{(3)} $ is BRST closed, one can easily derive the condition \cite{membr}
\begin{equation}\label{4f}
D_{(A} A_{BCD)} = \G^{M}_{(AB} A_{CD)M} \, ,
\end{equation}
where $A_{AB\,M}$ is an auxiliary superfield.\footnote{These equations are readily solved by imposing the Wess-Zumino gauge
$\theta^{A} A_{ABC}(x,\theta)=0$.} This form of the unintegrated vertex operator is related to the integrated one for the supermembrane \cite{membr}
$V^{(0)}_{membrane}=\int d^{3}x\, \mathcal{V}^{(0)}_{membrane}$ \,, where
\be
\mathcal{V}_{membrane}^{(0)} = 
{\partial}_{0}Z^\Sigma \Big(
A_{\Sigma \Gamma}  {\partial}_{0}Z^\Gamma + 
C_\Sigma{}^{A} d_{A} +  
\Omega_{\Sigma B}{}^{A} w_{A} \l^{B} \Big)  + \nn
\ee
\be
+
\Big(
    A_{\Sigma\Gamma\Lambda} {\partial}_{0}Z^\Sigma + 
    C_{\Gamma\Lambda}{}^{A} d_{A} +
    \Omega_{\Sigma\Lambda B}{}^{A} w_{A} \l^{B} 
\Big) \partial_{I}Z^\Gamma \partial_{J}Z^\Lambda \epsilon^{IJ} + \nn
\ee
\be
+
A_{\Sigma\Gamma\Lambda\Xi} 
\partial_{I}Z^\Sigma \partial_{J}Z^\Gamma 
\partial_{K}Z^\Lambda \partial_{L}Z^\Xi
\epsilon^{IJ}\epsilon^{KL} 
+ Y_{\Sigma B}{}^{A}\partial_{I}Z^\Sigma w_{A} \partial_J 
\l^{B} \epsilon^{IJ} \,. \label{membVop}
\ee
By setting to zero the transverse fluctuations of the 
membrane $\p_{I}Z^{\Sigma} =0$, one reduces (\ref{membVop}) to the first line, which is indeed of the same form as the vertex given in (\ref{VO}). 

\subsection{Coupling of the action to a background} 

Recall that in curved space we denote by capital greek letters $\Omega,  \Sigma, \dots$ superspace curved indices in 11d and by latin capital letters the tangent space indices. So $P_{M}$ and $d_{A}$ are the supersymmetric conjugate momenta to $Z^{\Omega}$. The superparticle action can be written as 
\begin{eqnarray}\label{curved} 
S_{curved} =\int d\tau\Big(  
 P_{M} E_{\Omega}{}^{M} \partial Z^{\Omega} +  
d_{A} E_{\Omega}{}^{A} \partial Z^{\Omega} - \frac{1}{2} \eta^{MN}  
P_{M} P_{N} +  
\end{eqnarray} 
$$ 
+w_{A} \Big( \partial \lambda^{A} + \partial Z^{\Sigma} \Omega_{\Sigma, [MN]} (\Gamma^{MN})_{A}^{B} \lambda_{B}\Big) \Big)\,.
$$ 
As always the connection couples to the ghost with a spin coupling. 
Clearly the equations of motion for the fields $x^{M}, \t^{A}, p_{A}$ and $P_{M}$ are changed.
 
In addition, the action (\ref{curved}) is no longer BRST invariant. This reflects itself in the lack of nilpotency of the BRST operator  
\begin{eqnarray} 
Q = \lambda^{A} \Big( d_{A} + E_{A \Sigma} \partial Z^{\Sigma} \Big) 
\end{eqnarray} 
in curved space. The non-invariance of the action implies that  
$\dot Q \neq 0$. Imposing 
 \bea\label{eqnarray}\label{nn} 
\dot Q = 0\,, ~~~~~~~~~ \{ Q, Q\} = 0  
\eea 
and using the pure spinor conditions, 
one ends up with the usual supergravity constraints on the  
supertorsion $T^{~~~~M}_{NR}$ and field strenght $G_{MNRS}$: 
\bea\label{con} 
T^{~~~~A}_{BC} = T_{A M}^{~~~N} =0\,, ~~~~ T_{AB}^{M} = \left(\Gamma^{M}\right)_{AB}
\,, 
\eea 
$$ 
G_{ABCD} =G_{ABCM}=0\,, ~~~~G_{ABMN} =\left(\Gamma_{MN}\right)_{AB}\,. 
$$ 
For later use we write down another version of the vertex operators (\ref{VO}) and (\ref{vopI}),
\begin{eqnarray}\nonumber 
V &=& \int d\tau\,  \left(\Pi_{M} + E_M{}^A d_A +  
H_{M}^{[PQ]} N_{PQ}  \right)\, g^{MN}\, \left(\Pi_{N} + E_{N}{}^{B} d_{B} +  
H_{N}^{[PQ]} N_{PQ}\right)\, e^{ik\cdot X}\\ 
U^{(1)}&=&  \left(\Pi_M + E_M{}^{A} d_{A} +  
H_M^{[NR]} N_{NR} \right)\,  E_B{}^{M}\lambda^{B}\, e^{ik\cdot X}
\label{Uone} \, ,
\end{eqnarray}
obtained after integrating out $P^{M}$ in (\ref{curved}). Notice that 
the integrated vertex operator, expressed in this form, 
resembles the integrated vertex of the closed superstring.


\section{Tree-level amplitudes} 

\subsection{Path integral definition for the tree-level measure}

It is convenient to rewrite (\ref{normtree}) in a functional integral 
form for later comparison with one-loop amplitudes. In order 
to be able to integrate over the 23 independent pure spinor components, 
we have to define a suitable measure ${\cal D}\l$ which respects the pure spinor constraint and by gauge fixing the 
zero modes of $\theta^{A}$ and $\l^{A}$ (at three level and for 
flat worldvolume with marked points, there are 
no zero mode for $w_{A}$ and $d_{A}$). So we introduce the Lagrange multiplier $\chi_{I}$ and its BRST partner $\eta_{I}$, where 
$I=1,\dots,23$, as well as 
their respective conjugate momenta $\bar\chi^{I}$ and 
$\bar\eta^{I}$. Now the BRST operator $Q$ changes to $Q_{new} =Q + \chi_{I} \bar \eta^{I}$. We also introduce the constant 
gauge fixing parameters $C^{I}_{A}$ with the help of which  
the tree level functional integral becomes 
\bea\label{treefunc}
\langle \prod_{i} U^{(n_{i})}_{i}(\l,X,\t) \rangle_{g=0} &=& 
\int d^{11}X d^{11} P 
d^{32}\t  d^{23} \chi d^{23} \eta [d\l^{A_{1}} \dots d\l^{A_{23}}] \\
&&
(\epsilon_{32} \mathcal{T}^{-1})_{[A_{1} \dots A_{23}]}^{(B_{1} \dots B_{7})} 
\prod_{j=1}^{7}
\frac{\partial}{\partial \l^{B_{j}}} 
\left(e^{[Q_{new}, 
\eta_{I} C^{I}_{A} \l^{A}]} \right) \prod_{i} U^{(n_{i})}_{i}(\l,X,\t) \,,
\nonumber
\eea
where $\sum_{i} n_{i} = 7$ and 
$(\epsilon_{32}\mathcal{T}^{-1})_{[A_{1} \dots A_{23}]}^{(B_{1} \dots B_{7})}$ is a unique tensor constructed in terms of $SO(10,1)$ Lorentz invariants. Integrating over $\chi_{I}$ and 
$\eta^{I}$, one can replace the exponential with 
the picture lowering operator $\prod_{I} C^{I}_{A} \t^{A} \delta(C^{I}_{A} \l^{A})$. 
This expression is manifestly independent of $C^{I}_{A}$ since 
the vertex operators $U^{(n_{i})}$, as well as the measure, are BRST invariant. In the following section we will see that 
the measure ${\cal D}\l$ is indeed consistent with the pure spinor 
constraint. Computing the derivatives in the above expression,
we end up with
\bea\label{treefuncII}
\langle \prod_{i} U^{(n_{i})}_{i}(\l,X,\t) \rangle_{g=0} &=&
\int d^{11}X d^{11} P
d^{32}\t  d^{23} \chi d^{23} \eta [d\l^{A_{1}}\wedge \dots \wedge
d\l^{A_{23}}] \\
&&
(\epsilon_{32} \mathcal{T}^{-1})_{[A_{1} \dots A_{23}]}^{(B_{1} \dots B_{7})}
\prod_{j=1}^{7} (C^{I}_{B_{j}} \chi_{I})
\left(e^{[Q_{new},
\eta_{I} C^{I}_{A} \l^{A}]} \right) \prod_{i} U^{(n_{i})}_{i}(\l,X,\t) \,.
\nonumber
\eea
Since the amplitude is independent of the 
gauge fixing parameters $C^{I}_{A}$, an easy way to compute it 
is to average over those parameters with a Gaussian-type 
measure such that $\int {\cal D} C = 1$.  

\subsection{Measure of integration for the pure spinor $\lambda$}

In this section we construct the bosonic ghost measure $[{\cal D}\l] $. The measure $[{\cal 
  D}w]$ for the canonically conjugate momentum $w_B$ (recall that $[ \lambda^A, w_B] =\delta^A{}_B$) 
  will be constructed in section~\ref{section:measureN} when discussing one-loop amplitudes. 
  
The  pure spinor condition 
\begin{equation}\label{puredef} 
\lambda \Gamma^M \lambda = 0 
\end{equation} 
is a set of first class constraints and for constructing the measure of integration we use the same method as in \cite{Berkovits:2004px}. Namely, from the zero mode prescription at tree-level 
(\ref{normtree}) we construct a tensor  ${\cal T}^{[A_1\cdots 
    A_9]}_{((B_1\cdots B_7))}$, antisymmetric in the $A_{i=1,\dots, 9}$ indices and 
traceless symetric in $B_{i=1,\dots,7}$, defined by
\begin{eqnarray}  
1&=&\langle (\lambda \Gamma^{m_1}\theta) \cdots (\lambda \Gamma^{m_7}\theta)\, (\theta \Gamma_{m_1\cdots m_7}\theta) \rangle\nonumber\\ 
 &=:&\lambda^{B_1}\cdots \lambda^{B_7}\,\theta^{A_1}\cdots \theta^{A_9} \, {\cal T}_{((B_1\cdots B_7))[A_1\cdots A_9]} 
\end{eqnarray}   
and write  
\begin{eqnarray}\label{dLambda} 
[d^{23}\lambda]^{[\alpha_1\cdots \alpha_{23}]}&:=& d\lambda^{\alpha_1} 
\wedge\cdots \wedge d\lambda^{\alpha_{23}} 
 = [{\cal D}\lambda]_{+16} \, (\epsilon_{32}{\cal T})^{[\alpha_1\cdots 
    \alpha_{23}]}_{((\beta_1\cdots \beta_7))}\, \lambda^{\beta_1} \cdots 
\lambda^{\beta_7} 
\end{eqnarray} 
with $\epsilon_{32}$ the totally antisymmetric tensor $\epsilon^{\alpha_1\cdots 
  \alpha_{32}}$ and   $[{\cal D}\lambda]_{+16}$ as scalar measure factor with 
ghost-charge $+16$.  

The inverse tensor $\mathcal{T}^{-1}$ is given by
\begin{eqnarray}\label{inverseT}
\left(\mathcal{T}^{-1}\right)^{((\alpha_{1}\cdots \alpha_{7}))}_{[\rho_{1}\cdots \rho_{9}]}&=&
 \langle \lambda^{\alpha_{1}}\cdots \lambda^{\alpha_{7}}\, \theta_{\rho_{1}}\cdots \theta_{\rho_{9}}\rangle\\
\nonumber &=&\Pi^{((\alpha_{1}\cdots \alpha_{7}))}_{\beta_{1}\cdots \beta_{7}}\, 
(\Gamma^{M_{1}})^{\beta_{1}}{}_{[\rho_{1}}\cdots (\Gamma^{M_{7}})^{\beta_{7}}{}_{\rho_{
7}]}\, (\Gamma_{M_{1}\cdots M_{7}} )_{\rho_{8}\rho_{9}]} \, ,
\end{eqnarray}
where $\Pi^{((\alpha_{1}\cdots \alpha_{7}))}_{\beta_{1}\cdots \beta_{7}}$ is the projector on symmetric $\Gamma_{M}$-traceless
7-tensors: 
\begin{eqnarray}
\Pi^{((\alpha_{1}\cdots \alpha_{7}))}_{\beta_{1}\cdots \beta_{7}}&=&
\delta^{(\alpha_{1}}{}_{\beta_{1}}\cdots 
\delta^{\alpha_{7})}{}_{\beta_{7}} - \frac{21}{74}\,(\Gamma^{M_{1}})^{(\alpha_{1}\alpha_{2}}
  \delta^{\alpha_{3}}{}_{(\beta_{1}}\cdots 
\delta^{\alpha_{7})}{}_{\beta_{5}}(\Gamma_{M_{1}})_{\beta_{6}\beta_{7})}\\
\nonumber&+&\frac{21}{592}\, (\Gamma^{M_{1}})^{(\alpha_{1}\alpha_{2}}  
(\Gamma^{M_{2}})^{\alpha_{3}\alpha_{4}}
  \delta^{\alpha_{5}}{}_{(\beta_{1}}\cdots 
\delta^{\alpha_{7})}{}_{\beta_{3}} (\Gamma_{M_{2}})_{\beta_{4}\beta_{5})}
(\Gamma_{M_{1}})_{\beta_{6}\beta_{7})}\\
\nonumber&-&\frac{21}{20128}\, (\Gamma^{M_{1}})^{(\alpha_{1}\alpha_{2}}  
\cdots (\Gamma^{M_{3}})^{\alpha_{5}\alpha_{6}}
\delta^{\alpha_{7})}{}_{(\beta_{1}} (\Gamma_{M_{3}})_{\beta_{2}\beta_{3}}\cdots
(\Gamma_{M_{1}})_{\beta_{6}\beta_{7})} \,.
\end{eqnarray}

Before concluding this subsection we make a few remarks to stress the differences between the eleven-dimensional case studied in the present paper and the ten-dimensional setup. 

We can parametrize the pure spinor $\lambda^A$ by first splitting the 
{\bf 32} of SO(11) as ${\bf 16}\oplus {\bf 16}$ of SO(10), i.e. $\lambda^A = 
(\lambda_L^\alpha, \lambda_R^\alpha)$,  and then decomposing each SO(10) spinor 
into U(5) representations ${\bf 1}\oplus {\bf 5}\oplus {\bf\bar{10}}$ and 
${\bf\bar{ 1}}\oplus {\bf\bar{ 5}}\oplus {\bf 10}$ as  
$$ 
\lambda_L^\alpha= (\lambda_+,\lambda_a,\lambda^{ab}),\quad \lambda_R^\alpha= 
(\lambda_-,\lambda^a,\lambda_{ab}), \qquad a=1,\cdots,5  
$$ 
In this language the pure spinor constraint reads 
\begin{eqnarray} 
\lambda_+\lambda_- + \lambda_a \lambda^a + \frac{1}{2} \lambda_{ab} \lambda^{ab} 
&=&0\nn\\  
\lambda_+ \lambda_a + \frac{1}{8} \epsilon_{abcde} \lambda^{bc} \lambda^{de} 
&=&- \lambda^b \lambda_{ab} \nn\\  
\lambda_- \lambda^a + \frac{1}{8} \epsilon^{abcde} \lambda_{bc} \lambda_{de} 
&=&- \lambda_b \lambda^{ab} \label{UfiveOne}
\end{eqnarray} 
where the first equation comes from $\lambda \Gamma^{11}\lambda =0$ and the 
other two from $\lambda_L \Gamma^m \lambda_L + \lambda_R \Gamma^m \lambda_R = 
0$. We solve these equations as   
\begin{eqnarray}\label{UfiveTwo} 
5! \lambda^{[a} \lambda^{bc} \lambda^{de]} &=&8 \epsilon^{abcde}\lambda_+ 
\left(\lambda_+\lambda_-+ \frac{1}{2} \lambda_{ab} \lambda^{ab}\right) \nn\\  
5! \lambda_{[a} \lambda_{bc} \lambda_{de]} &=&8 \epsilon_{abcde}\lambda_- 
\left(\lambda_+\lambda_-+ \frac{1}{2} \lambda_{ab} \lambda^{ab}\right) .  
\end{eqnarray} 
This determines completely all components of $\lambda_a$ and $\lambda^a$. The solution is very similar to the one of the ten-dimensional 
pure spinor constraint, where the {\bf 5} is solved in terms of the ${\bf\bar{ 
    10}}$ of U(5) as $8\lambda_+ \lambda_a = 
-\epsilon_{abcde}\lambda^{bc}\lambda^{de}$ (see \cite{membr}). But there is a 
crucial difference since the {\bf 5} of $\lambda_L$ is solved in terms of the 
{\bf 1} and {\bf 10} of $\lambda_L$ and $\lambda_R$. This gives a solution with 
$2\times (1+10)=22$ complex parameters. The SU(5) decomposition of an SO(11) pure spinor has an extra complex parameter $\rho$, which leaves equations 
(\ref{UfiveOne}) and (\ref{UfiveTwo}) invariant under the rescaling 
\begin{equation}\label{Scale} 
(\lambda_L, \lambda_R) \to (\rho \, \lambda_L, \rho^{-1} \, \lambda_R) \,. 
\end{equation} 
This symmetry is related to 
the constraint $\lambda \Gamma^{11}\lambda=0$ (i.e. 
$\l_{L,\a} \l^{\a}_{R} =0$; this condition can be solved by a generic choice of 
$\l_{\a,L}$ and $\l^{\a}_{R}$ up to a scale gauge symmetry), which 
states that an eleven dimensional pure spinor is not just the square of two ten 
dimensional pure spinors.  
It is now clear that the measure on $\lambda^A$ can also be written in the following form:  
\begin{equation}\label{MeasureI} 
d^{23}\lambda^A = d\rho\wedge \left(d\lambda_+ \wedge d\lambda^{ab}\right)\wedge 
\left(d\lambda_- \wedge d\lambda_{ab}\right) \,.  
\end{equation}

\subsection{Tree-level action from scattering amplitudes}

In this subsection we will reproduce the effective action for eleven dimensional supergravity up to four-fermi terms from tree-level amplitude
computations. The supergravity action up to this order in the fermion fields reads \cite{CJS}
\begin{equation}\label{CJS}
S = \frac{1}{2\ell_{P}^{9}}\int d^{11}x \left[e\,\mathcal{R}+ e\,\bar\Psi_{M}\Gamma^{MNP}\hat
D_{N}\Psi_{P} + \frac{1}{2}G_{4}\wedge * G_{4}+\frac{1}{6} C_{3}\wedge G_{4}\wedge G_{4}\right] \, ,
\end{equation}
where $\hat D_{M} = \partial_{M} +\frac{1}{4} \omega_{M}^{rs}\Gamma^{rs}+ \mathcal{T}_{M}{}^{r_{1}\cdots r_{4}} G_{r_{1}\cdots r_{4}}$
is the supercovariant derivative (again up to fermion bilinears), involving the
$G_{4}$ field strength. The
 tensor  $\mathcal{T}_{M}{}^{r_{1}\cdots r_{4}}$ is
defined in~(\ref{tensorCJST}).

The zero momentum cohomology of $Q$ contains a non zero $\lambda$-ghost
number 3 vertex operator $U^{(3)}$ whose components are comprised of the
supergravity fields and their derivatives and a $\lambda$-ghost number 4 vertex
operator $U^{(4)}$ containing the antifields. The relevant components of
$U^{(3)}$ are \cite{membr}:
\begin{eqnarray}
\nonumber
U^{(3)}_{C_{3}}&=&(\lambda\Gamma^{M}\theta)(\lambda\Gamma^{N}\theta)(\lambda\Gamma^{K}\theta)\,
C_{MNK}\, e^{ik\cdot X}\\
\nonumber 
U^{(3)}_{g}&=&(\lambda\Gamma^{(M}\theta)(\lambda\Gamma^{N)K}\theta)(\lambda\Gamma_{K}\theta)\,
g_{MN}\, e^{ik\cdot X}\\
U^{(3)}_{\Psi}&=& (\lambda\Gamma^{M}\theta)\left[(\lambda\Gamma^{N}\theta)(\lambda\Gamma^{P}\theta)
(\theta\Gamma^{NP}\Psi_{M})\right.\\
\nonumber&-&\left.(\lambda\Gamma^{NP}\theta)(\lambda\Gamma^{N}\theta)(\theta\Gamma^{P}\Psi_{M})\right]
\, e^{ik\cdot X}\, .
\end{eqnarray}
Derivatives of the supergravity fields appear at higher orders in the $\theta$
expansion of $U^{(3)}$. The $U^{(4)}$ vertex operator has a similar expression
in terms of the antifields. However, it does not enter the definition of tree-level superparticle amplitudes as it is intrinsically
related to the supermembrane. We will comment more on $U^{(4)}$ below.

For performing amplitude
computations we also need the part of the $U^{(1)}$ vertex operator~(\ref{vopI})
which contains the gravitino and the 3-form field strength:
\begin{eqnarray}
U^{(1)}_{\Psi}&=& (\lambda \Gamma_{N}\theta)(\theta \Gamma_{M}\Psi^{N})\,
P^{M}\, e^{ik\cdot X}\\
U^{(1)}_{G_{4}}&=& (\lambda \Gamma_{N}\theta)(\theta \mathcal{U}^{N|r_{1}\cdots r_{4}M}\theta)
\, G_{r_{1}\cdots r_{4}}\, P^{M}\, e^{ik\cdot X} \, .
\end{eqnarray}
The tensor $\mathcal{U}^{N|r_{1}\cdots r_{4}M}$ is defined in~(\ref{tensorCJSU}).
Because $U^{(3)}_{C_3,g,\Psi}$ are physical vertex operators,  $\{Q,U^{(3)}\}=0$ implies the equations of motion of eleven 
dimensional supergravity \cite{membr}. This means that all terms containing two fields in the action~(\ref{CJS})
are derivable from  the amplitude
$\langle U^{(3)}|Q| U^{(3)}\rangle$. The cubic term in the supergravity action~(\ref{CJS})
is obtained from the amplitude $\langle U^{(3)} U^{(1)} U^{(3)}\rangle$.

For instance, we first consider the amplitude given by the insertion of two $U^{(3)}$ vertex operators for the $C_{3}$-field at positions $\tau_{1}$
 and $\tau_{2}$ and a $C_{3}$ vertex operator $U^{(1)}$ at position $\tau_{3}$:
\begin{eqnarray}
&&\nonumber\langle U^{(3)}_{C_{3}}(\tau_{1}) U^{(3)}_{C_{3}}(\tau_{2})
 U^{(1)}_{G_{4}}(\tau_{3})\rangle = C_{M_{1}\cdots M_{3}}\,C_{N_{1}\cdots N_{3}}\, G_{r_{1}\cdots r_{4}}\\
\nonumber&\times& \langle(\lambda\Gamma^{M_{1}}\theta)\cdots
(\lambda\Gamma^{M_{3}}\theta)(\lambda\Gamma^{N_{1}}\theta)\cdots
(\lambda\Gamma^{N_{3}}\theta)
 (\lambda \Gamma^{N}\theta)(\theta \mathcal{U}^{N|r_{1}\cdots r_{4}M}\theta)\,
P^{M} \prod_{i=1}^{3}e^{ik^{(i)}\cdot X}\rangle \, .
\end{eqnarray}
Using that on shell $(k^{(i)})^{2}=k^{(i)}\cdot k^{(j)}=0$ for $i,j=1,2,3$,
$P^{M}=\dot X^{M}$ and also the
tree-level correlator of the world-line formalism \cite{Strassler:1992zr}: $$
\langle X^{M}(\tau_{1}) X^{N}(\tau_{2})\rangle_{tree}= \eta^{MN}\left(\frac{|\tau_{1}-\tau_{2}|}{2}+a +
b\tau_{2}\right)
\, , $$ 
where $a$ and $b$ are arbitrary constants not entering in any physical amplitude,
the contraction with the plane-wave from a $U^{(3)}$ vertex operator 
 gives $\langle P^{M}(\tau_{3})\exp(i k^{(i)}\cdot X(\tau_{i}))\rangle= i (k^{(i)})^{M}\, \text{sign}(\tau_{13})$. So assuming that $\tau_{1}<\tau_{3}<\tau_{2}$ we obtain:
 \begin{eqnarray}
&& \langle U^{(3)}_{C_{3}}(\tau_{1}) U^{(3)}_{C_{3}}(\tau_{2}) U^{(1)}_{G_{4}}(\tau_{3})\rangle =  \frac{1}{288} C_{M_{1}\cdots M_{3}}\,C_{M_{4}\cdots M_{6}}\, 
G_{r_{1}\cdots r_{4}} [(k^{(2)})_{L}-(k^{(1)})_{L}]\\
\nonumber&\times&\langle(\lambda\Gamma^{M_{1}}\theta)\cdots
(\lambda\Gamma^{M_{7}}\theta)\,  \left[5 \eta^{M_{7}[L}(\theta  \Gamma^{r_{1}\cdots r_{4}]}\theta)
+8(\theta\Gamma^{M_{7}[r_{1}\cdots r_{3}}\theta)\eta^{r_{4}]L}\right]\rangle\ .
 \end{eqnarray}
Because of the relation
$$
\langle(\lambda \Gamma^{M_{1}}\theta)\cdots (\lambda\Gamma^{M_{7}}\theta)(\theta\Gamma^{M_{8}\cdots M_{11}}\theta) \rangle \propto \epsilon_{11}^{M_{1}\cdots M_{11}} \, ,
$$
with the proportionality constant fixed by the normalization~(\ref{normtree}),
this amplitude reproduces the Chern-Simons coupling of the supergravity action~(\ref{CJS})
\begin{equation}
\langle U^{(3)}_{C_{3}}U^{(3)}_{C_{3}} U^{(1)}_{G_{4}}\rangle =\epsilon_{11}{}^{M_{1}\cdots M_{11}}\, C_{M_{1}\cdots M_{3}}G_{M_{4}\cdots M_{7}} G_{M_{8}\cdots M_{11}} \, .
\end{equation}
The other coupling $\Psi_{M}\Gamma^{MNP}\mathcal{T}_{N}\cdot G_{4}\Psi_{P}$ from the supercovariant derivative
can be derived in the same way from the amplitude $\langle U^{(3)}_{C_{3}}  U^{(3)}_{\Psi} U^{(1)}_{\Psi}\rangle$, 
which is the only one compatible with the tree level normalisation $\langle \lambda^{7}\theta^{9}\rangle$. 

We can therefore summarize the two-derivative non-linear eleven dimensional supergravity action in the following way:
\begin{equation} \label{defD}
S_{11d} = \langle U^{(3)} Q U^{(3)}\rangle + \langle U^{(3)} [U^{(1)}, U^{(3)}] \rangle \, ,
\end{equation}
where $[U^{(1)}, U^{(3)}]$ means using the canonical bracket for the on-shell
fields $P^{M}$ and $X_{N}$ appearing in $U^{(1)}$ and $U^{(3)}$. The symbol
$\langle \cdots \rangle$ denotes the correlator for zero modes using the
measure~(\ref{normtree}). The action  (\ref{defD}) varies under the gauge
transformation $\delta U^{(3)} = \{Q + U^{(1)}, \Omega^{(2)}\}$ into $\langle
U^{(3)}[U^{(1)},\{U^{(1)}, \Omega^{(2)}\}]\rangle$. The first term in the
transformation, $\{Q, \Omega^{(2)}\}$,  generates gauge symmetries on the
physical fields, whereas the second term,
$\{U^{(1)}_{\Phi},\Omega^{(2)}\}= \delta U^{(3)}_{\delta \Phi}$, generates
supersymmetry transformations $\delta\Phi$ on the physical fields. The nonzero
variation of the action is due to neglecting\footnote{The reason we have not written out these terms
is only aesthetical, not conceptual. They arise as contact terms
of the four-point amplitude containing an integrated vertex operator.} the four
fermion terms in (\ref{defD}) necessary for supersymmetry invariance of (\ref{CJS}).\footnote{Recall that they are taken care of by
completing the spin connection and four-form with fermion bilinears to get a
supercovariant derivative
acting on the fermions \cite{CJS}.}


\section{Definition of the one-loop amplitude} \label{section:oneL}

Using a world-line path integral formulation of a loop amplitude, it is necessary to integrate over the Schwinger 
parameters, the loop momenta, the 32 fermionic variables $\theta^A$, the 32 
supersymmetric Green-Schwarz constraints $d_A$ (representing the conjugate momenta $p_A$), the pure spinors 
$\lambda^A$ and their conjugate ghosts $w_A$.  
As explained in \cite{Berkovits:2004px} it  
is as well necessary to insert picture raising operators $Z_B$ and $Z_J$
\bea
Z_B &=& B_{MN} (\l\Gamma^{MN}d) \,\delta\Big(B_{MN} N^{MN}\Big) \\
Z_J &=& \l d \,\delta(J)
\eea
and picture lowering operators
\be
Y_C = C_{A}\theta^{A} \delta\Big( C_{A} \l^{A}\Big)
\ee 
for absorbing the zero modes of $\lambda^A$ and $w_A$. As usual, at 
one loop translation invariance requires to have at least one unintegrated vertex operator. Summarizing, the amplitude is defined as: 
\be 
\label{oneL} 
\mathcal{A} = 
\int \frac{d\tau}{\tau} 
\Big \langle  
b_{B} \,  
Z_{J} \, \prod_{P=2}^{22} Z_{B}  \, \prod_{I=1}^{23} Y_{C_{I}} 
 \, U^{(1)}(\tau_{1})  
 \prod_{T=2}^{N} \int_0^{\tau} d\tau^{T} {\cal V}(\tau^{T}) 
\Big\rangle \, ,
\ee 
where $U^{(1)}$ is the unintegrated vertex operator defined as $\{Q, {\cal
V}(\tau)\} =: 
\partial_\tau U^{(1)}$. Notice that the vertex operator $U^{(1)}$ has
ghost-number 1 and contains $d_A$ terms (see (\ref{Uone})). \footnote{We
already pointed out that the tree-level amplitudes for the superparticle do not involve the
ghost number 4 vertex operator $U^{(4)}$. By unitary arguments, this vertex does not enter the definition of superparticle loop amplitudes either.}

Before specifying the measure of integration over the $\lambda^\alpha$ and 
$w_\alpha$, we first stress that this amplitude cannot be just a trivial lift of 
the one-loop superstring prescription of Berkovits, since the number of 
components of $\lambda^\alpha$ is 23 which is with one more than for the left/right doublet of the ten-dimensional pure spinors  
$(\lambda_L^\alpha,\lambda_R^\alpha)$.  

Let us also check that the zero mode counting for $\theta^A$, $d_A$, and $\lambda_A$ works: 

\begin{itemize}

\item[$\triangleright$] We start by enumerating the $d_\alpha$ zero modes. At
one-loop 32 have to be soaken up.
Clearly one gets 21 from $Z_{B}$, 1 from $Z_{J}$, 1 from $U^{(1)}$ and at most
$2(N-1)-M$ from the integrated vertex operators ${\cal V}$ (notice that the
${\cal V}$'s depend on the quadratic combinations: $dd$, $dN$ or $NN$), where
$M$ is the number of $N^{NM}$ zero modes. So far, the counting of $d_{\a}$ zero
modes gives $32 - 21 - 1 - 1 +2 - 2N + M = 11- 2N + M$. Therefore for the
four-point amplitude, $N=4$, we have to extract $4 \,d$'s from $b_{B}$. Since
the latter come together with a $\delta'(B_{MN} N^{MN})$, which counts as $-1$
$N^{NM}$ zero mode, we also have to pick $M=1$ to saturate it. Morever, the above
counting shows that the first non-vanishing one-loop amplitude is 
with four external states. 

\item[$\triangleright$] Now we count the $\theta$ zero modes. There are 32 of them to 
  integrate over; the picture lowering operators $Y_C$ contain 23 so 9 have to 
  arise from the vertex operators. For the case $N=4$ and $M=1$, we already get 
  5 $\theta$'s from the integrated vertex operators. So $U^{(1)}$ has to 
  contribute 4 $\theta$ zero modes.

\item[$\triangleright$] The $\lambda$-ghost counting goes as follows: We get $+21$ zero modes from $Z_{B}$, $+1$ from 
$Z_{J}$, $+1$ from $U^{(1)}$ and $-23$ from $Y_C$, for a  total 
of  
\begin{equation}\label{ghostcharge} 
21+1  -23 + 1=0\ . 
\end{equation} 

\end{itemize}

\subsection{Measure for the conjugate momentum $w_{A}$}\label{section:measureN}

The constraints (\ref{puredef}) allow the gauge symmetry $\delta_\Lambda w_{A} 
= \Lambda_{M}  \Gamma^{M}_{AB} \l^{B}$. The only bilinears, $(w\G^{M_1\cdots M_r}\l)$, invariant under it are the Lorentz generators $N^{MN}=(w \G^{MN} \l)/2$ and the ghost-number operator $J=(w \lambda)$. The reason is that the variation 
$\delta_\Lambda(w\G^{M_1\cdots M_r}\l)=(\delta_\Lambda w\G^{M_1\cdots M_r}\l)  = 
\Lambda_P \, (\lambda \G^{PM_1\cdots M_r}\l) + r \Lambda^{[M_1} \, (\lambda 
  \G^{M_2\cdots M_r]}\l)$ vanishes only for $r=0$ and $r=2$ because of the commuting nature of $\lambda^A$. Generally the vertex operators are also  
 invariant under $\delta_\Lambda$ and depend on $N^{[MN]}:=\frac{1}{2}\, (w\Gamma^{MN}\lambda)$ and $J=(w\lambda)$. To construct the measure for integration over $w_A$ we use the change of variables  
\begin{equation} 
32\, w_A \, \lambda^B = \delta_A{}^B \, J + \frac{1}{2!}\, 
\left(\Gamma_{MN}\right)_A{}^B \, N^{MN}  
\end{equation} 
and expand $\wedge^{23} dw_A$ as 
\begin{equation}\label{dNone}
\left(\wedge^{23} dw_A \right) \lambda^{B_1} \cdots \lambda^{B_{23}}= 
\mathcal{T}_1^{m_1n_1\cdots m_{22}n_{22}}  \left(\wedge^{22}dN^{[m_in_i]} \wedge 
dJ\right) + \mathcal{T}_2^{m_1n_1\cdots m_{23}n_{23}} 
\left(\wedge^{23}dN^{[m_in_i]}\right) .
\end{equation}
Doing so we have obtained two contributions to the measure. The tensors 
$\mathcal{T}_{I=1,2}^{m_1\cdots }$ will be determined later. 

The measure factors 
in (\ref{dNone}) satisfy respectively  
\begin{eqnarray}\label{ConstraintI}  
(\lambda \Gamma_{m_1})_A\, \left(\wedge^{22}dN^{[m_in_i]}\wedge dJ\right)&=&0\\ 
\label{ConstraintII} 
(\lambda \Gamma_{m_1})_A\, \left(\wedge^{23}dN^{[m_in_i]}\right)&=&(\lambda \Gamma^{n_1})_A\, \left(\wedge^{22}dN^{[m_in_i]}\wedge dJ\right)\ . 
\end{eqnarray} 
The first equation follows from the relation
\begin{equation} 
N^{MN} \lambda^A \, (\Gamma_M)_{AB} - \frac{1}{2} J \lambda^A \, (\Gamma^N)_{AB} =0 \, ,
\end{equation} 
valid at the classical level. The second one can be checked directly by using the Fierz identities~(\ref{GammaSix}), (\ref{Idtwo}), (\ref{Idthree}) and~(\ref{Idfour}). Equation (\ref{ConstraintII}) shows that the second contribution in (\ref{dNone}) is not 
independant of the first one. Hence there is a single measure factor for 
$\wedge^{23}dw_A$ compatible with the pure spinor constraint (\ref{puredef}). 

Counting the number of available Lorentz indices and having at most 23 
$\lambda^A$ we find that the measure can be decomposed in the following two 
products of $\lambda$-bilinears (see Appendix~\ref{section:Fierz} for details about Fierz identities in eleven dimensions):\footnote{In the following the antisymmetric product of $n$ Gamma-matrices $\Gamma_{M_1\cdots M_n}$ is abreviated as $\Gamma_{[n]}$.}  
\begin{equation}\label{dNmeasure} 
\wedge^{22}dN^{[m_in_i]}\wedge dJ=[\mathcal{D}w]_{-16} \, (\lambda\Gamma_{[5]}\lambda)^4(\lambda\Gamma_{[6]}\lambda)^4 
+ [\mathcal{D}w]_{-20} (\lambda\Gamma_{[5]}\lambda)^8(\lambda\Gamma_{[2]}\lambda)^2 \, ,
\end{equation}  
where $[\mathcal{D}N]_{g=-16,-20}$ are scalar measure factors of ghost-number 
$-16$ and $-20$. The constraint (\ref{ConstraintI}) and the identities in Appendix~\ref{section:Fierz} forbid the ghost-number $-20$ term, but allow the other one with 
ghost-number $-16$:  
\begin{eqnarray}\label{dN}  
\wedge^{22}dN^{[m_in_i]}\wedge dJ&=&[\mathcal{D}w]_{-16} \times\\ 
&\times&\left[ 
  (\lambda\Gamma^{m_1n_1m_2m_3m_4}\lambda)(\lambda\Gamma^{m_5n_5n_2m_6m_7}\lambda)(\lambda\Gamma^{m_8n_8n_3n_6m_9}\lambda)(\lambda\Gamma^{m_{10}n_{10}n_4n_7n_9}\lambda)\right. \nonumber\\  
&\times& 
  (\lambda\Gamma^{m_{11}n_{11}m_{12}m_{13}m_{14}m_{21}}\lambda)(\lambda\Gamma^{m_{15}n_{15}n_{12}m_{16}m_{17}n_{21}}\lambda)(\lambda\Gamma^{m_{18}n_{18}n_{13}n_{16}m_{19}m_{22}}\lambda) \nonumber\\  
&\times&\left.(\lambda\Gamma^{m_{20}n_{20}n_{14}n_{17}n_{19}n_{22}}\lambda)+perms\right] \,.
\nonumber  
\end{eqnarray} 
Remarking that $(\lambda \Gamma_{[6]} \lambda)= \epsilon_{11}(\lambda\Gamma_{[5]}\lambda)$ we can interpret this product as the left-movers times the right-movers measure factor constructed for the superstring in \cite{Berkovits:2004px}. 
For later use, we also note that the Ward identity~(\ref{ConstraintII}) allows an alternative form for the measure factor:
\begin{eqnarray}\label{dNbis}
\wedge^{23}dN^{[m_in_i]}&=& [\mathcal{D}w]_{-16}\times
\left[  (\lambda\Gamma^{m_1n_1m_2m_3m_4}\lambda)(\lambda\Gamma^{m_5n_5n_2m_6m_7}\lambda)\right.\\
 &\times&(\lambda\Gamma^{m_8n_8n_3n_6m_9m_{23}}\lambda)(\lambda\Gamma^{m_{10}n_{10}n_4n_7n_9n_{23}}\lambda)\nonumber\\  
&\times&   (\lambda\Gamma^{m_{11}n_{11}m_{12}m_{13}m_{14}m_{21}}\lambda)(\lambda\Gamma^{m_{15}n_{15}n_{12}m_{16}m_{17}n_{21}}\lambda)(\lambda\Gamma^{m_{18}n_{18}n_{13}n_{16}m_{19}m_{22}}\lambda) \nonumber\\  
&\times&\left.(\lambda\Gamma^{m_{20}n_{20}n_{14}n_{17}n_{19}n_{22}}\lambda)+perms\right] \,.
\nonumber  
\end{eqnarray}

\subsection{A different representation of the  $b_B$ ghost} 

An important peculiarity of Berkovits' pure spinor cohomology is that the energy momentum tensor $T= P^{2}/2$ is not $Q$-exact. One can define 
a (bi-local) ghost $\widetilde b_{B}$ by requiring that \cite{Berkovits:2004px} 
\be \label{oneB} 
\Big\{Q, \widetilde b_{B}(u,z) \Big\} = Z_{B}(u) T(z) \, ,  
\ee 
 where  
  $$Z_{B} = B_{MN} (\l\Gamma^{MN}d) \delta\Big(B_{MN} N^{MN}\Big)\ .$$ 
  Notice that, since we are considering the superparticle, all the fields entering $Z_{B}$ are constant and therefore we do not need to specify the position. The same is also true for the picture raising operator $Y_{C_{I}}$ defined by $Y_C = C_{A}\theta^{A} \delta\Big( C_{A} \l^{A}\Big)$. Nevertheless, we will keep writing the position argument for generality as the manipulations we do in the rest of this subsection are valid for string theory as well.

Instead of working with a bi-local ghost $\widetilde b_{B}(u,z)$ we consider a 
local one $\widetilde b_{B}(u)$ defined up to exact terms  as 
\bea \label{oneD} 
\widetilde b_{B}(u,z) = \widetilde b_{B}(u) + \{Q, \int_{u}^{z} \Omega\}\,, 
 \eea 
where $\Omega(u,v)$ is a bi-local superfield. 
In addition, by recalling that $\p \Theta(x) = \delta(x)$, we find an 
alternative to the solution of \cite{Berkovits:2004px}:  
\bea\label{oneE} 
\{Q, \widetilde b_{B}(u) \} =  
B_{MN} \l\G^{MN} d \, \delta\Big(B_{MN} N^{MN} \Big) T(u)   =  
\{Q, \Theta\Big(B_{MN} N^{MN} \Big) T(u) \}  
\eea 
where we have used the BRST invariance of $T(u)$. Hence 
\bea\label{oneG} 
\widetilde b_{B}(u) = T(u)  \Theta\Big(B_{MN} N^{MN} \Big) + \{Q, \Omega\}\,, 
\eea 
where the second BRST-exact term is clearly unimportant.  
The same representation applies to the left and right $b_B^{L/R}$ ghosts for 
type II superstrings. 

Finally notice that, using an integral representation for the $\Theta$-function, the insertion of the $b_B^{L/R}$ in the one-loop superstring amplitude, can be rewritten as  
\bea\label{oneM} 
\int \mu(z, \bar z) b^{L}_{B} b^{R}_{B} =  
\int \mu(z,\bar z) \int dt ds  
\frac{e^{i  B_{mn} [(t+ s) (N^{L,mn}  +  N^{R,mn}) +  
(t-s) (N^{L,mn}  -  N^{R,mn})]} 
}{ 
(t + i \epsilon) ( s + i \epsilon) }\ .
\eea 
Clearly, the second term in the exponent, that involves the difference of the Lorentz generators, 
imposes the level matching condition.


\subsubsection{Action of the $b_B$ ghost on vertex operators} 
Let us consider the action of the $b_B$ ghost on a vertex operator $U^{(q+1)}(\tau)$ with ghost-number $(q+1)$:
\bea\label{oneI} 
\widetilde b_{B}(u) U^{(q+1)}(\tau) =  
\Theta\Big(B\cdot N\Big) T(u) U^{(q+1)}(\tau)\,.  
\eea  
Using the fact that for $q\geq 0$,  $U^{(q+1)} = \l^{A} \Phi^{(q)}_A$ and that  
$\{Q, G^{A}\} = \frac{P^{2}}{2} \l^A$, where  
$G^{A} = \G^{AB}_{M} P^{M} d_{B}$, we can rewrite (\ref{oneI})  as 
\bea\label{oneJ} 
&&\Theta\Big(B\cdot N\Big)T(u) U^{(q+1)}(\tau) =  
\Big\{Q, G^{A}\Big\} \Phi^{(q)}_A   \Theta\Big(B\cdot  N\Big)  \\ 
&=&\Big\{Q, G^{A}  \Phi^{(q)}_A \Theta\Big(B\cdot N\Big) \Big\}  
- G^{A} \l^{B} D_{B}\Phi^{(q)}_A \Theta\Big(B\cdot N\Big) - 
G^A \Phi^{(q)}_A\,  Z_B\,, \nonumber 
\eea 
Now we can use that $\{Q, H^{AB}\} = G^{A} \l^{B} + g^{((AB))}$, 
where $H^{AB}$  
is given by\footnote{It is easy to see that the structure of $H^{AB}$ is the following: \be H^{AB} = X^{AB[CD]} d_C d_D + Y^{AB}_{MNP} P^M N^{NP} + Z^{AB}_M P^M J \, , \nn \ee where \bea X^{AB[CD]} &=& {\bf x} C^{AB} C^{[CD]} + {\bf x'} \G^{AB}_{MNP} (\G^{MNP})^{[CD]} + {\bf x''} \G^{AB}_{MNPQ} (\G^{MNPQ})^{[CD]} \nn \\ Y^{AB}_{MNP} &=& {\bf y} \G^{AB}_{MNP} + {\bf y'} \G^{AB}_{N} \eta_{PM} \nn \\ Z^{AB}_M &=& z \G^{AB}_M \nn \eea and ${\bf x, x', x'', y, y', z}$ are unknown numerical constants. What is more involved is to calculate the numerical coefficients. One can do that by computing $\{ Q, H^{AB}\} - \l^A G^B$. Since this is equal to $g^{((AB))}$, i.e. it is symmetric and traceless, one finds conditions for the unknown constants from the requirement that the trace and antisymmetric part of this expression vanish. It is most easy to extract those conditions by contracting $\{ Q, H^{AB}\} - \l^A G^B$ with $\G^{M'}_{AB}, \G^{M'N'P'}_{AB}, \G^{M'N'P'Q'}_{AB}$ and $C_{AB}$. Two of the resulting four equations contain two independent structures each and so one obtains six equations which determine uniquely the six constants.}  
\bea\label{oneK} 
&&H^{AB} = \frac{1}{64}\, \left[ C^{AB} (d_{C} C^{CD} d_{D}) + \G^{AB}_{MNPQ} (d \G^{MNPQ} d) \nonumber \right. \\ &&~~~~~~~+ \left. \G^{AB}_{MNP}( d \G^{MNP} \, d +2\, N^{MN} P^{P} ) + 2\, \G^{AB}_{M} ( N^{MR} P_{R} + J P^{M}) \right] 
\eea 
and $g^{((AB))}$ is traceless and symmetric, to reabsorb part of the vertex into $Q$-exact terms: 
\bea\label{oneKb} 
\Theta\Big(B_{MN} N^{MN}\Big) T(u) U^{(q+1)}(\tau) &=&  - \left[G^A \Phi^{(q)}_A \, -H^{AB} D_A \Phi^{(q)}_B\right]\, Z_B+ \{Q, \Omega\}\\
&+&\left[ H^{AB} \l^{C}   
D_CD_B\Phi^{(q)}_A 
  \nonumber+ g^{((AB))} D_A \Phi^{(q)}_B\right] \Theta\Big(B\cdot N\Big)
\eea 
Then, making use of $\{Q, K^{ABC}\}= H^{AB} \l^C+ h_1^{((AB))C}+ h_2^{A((BC))}$, we rewrite the first term on the second line as before:
 \bea\label{oneKc} 
&&\Theta\Big(B_{MN} N^{MN}\Big) T(u) U^{(q+1)}(\tau) =  \{Q, \Omega\}\\
\nonumber& -& \left[G^A \Phi^{(q)}_A \, -H^{AB} D_A \Phi^{(q)}_B-K^{ABC} D_CD_B\Phi^{(q)}_A\right]\, Z_B\\
\nonumber&+&\left[ K^{AB} \l^{D}  D_D 
D_CD_B\Phi^{(q)}_A +g^{((AB))} D_A \Phi^{(q)}_B+h_i^{ABC} D_CD_B\Phi^{(q)}_A\right] \Theta\Big(B\cdot N\Big)\ .
\eea 
Finally, using that $\{Q, L^{ABCD}\}= \l^A K^{BCD} + k_1^{((AB))CD} + k_2^{A((BC)) D}+k_3^{AB((CD))}$, we obtain:
 \bea\label{oneKd} 
&&\Theta\Big(B_{MN} N^{MN}\Big) T(u) U^{(q+1)}(\tau) =  \{Q, \Omega\}\\
\nonumber& -& \left[G^A \Phi^{(q)}_A \, -H^{AB} D_A \Phi^{(q)}_B-K^{ABC} D_CD_B\Phi^{(q)}_A+L^{ABCD} D_D D_C D_B \Phi^{(q)}_A\right]\, Z_B\\
\nonumber&+&\left[ g^{((AB))} D_A \Phi^{(q)}_B+h_i^{ABC} D_CD_B\Phi^{(q)}_A+k_i^{ABCD} D_D D_C D_B \Phi^{(q)}_A\right] \Theta\Big(B\cdot N\Big)\ .
\eea 

However, due to $\{D_A,D_B\} = P_M \, (\G^M)_{AB}$ and $D_{(A} \Phi_{B)}^{(q)}\sim (\G^M)_{AB} \Phi_M^{(q)}$ all tensors appearing in the third line are $\G_{[1]}$-traceless and so the whole expression multiplying $\Theta(B\cdot N)$ vanishes. We can interpret the second line in (\ref{oneKd}) as the vertex operator $U^{(q+1)}$ as written in a different '$b$-picture':   
\begin{eqnarray}
\nonumber
V^{(q+1)}_{B}(\tau)&=& - \left[G^A \Phi^{(q)}_A \, -H^{AB} D_A \Phi^{(q)}_B-K^{ABC} D_CD_B\Phi^{(q)}_A+L^{ABCD} D_D D_C D_B \Phi^{(q)}_A\right]\, Z_B\\
&=& \hat V^{(q)}(\tau) \, Z_B \, .
\label{newvop} \end{eqnarray}  
Note that these manipulations do not change the ghost number, since $Z_B$ has ghost number 1 and $\hat V^{(q)}$ has ghost number $q$. Comparing with Berkovits' expression for the $b(z)$-ghost in section~4.2 of \cite{Berkovits:2004px}, the vertex operator $\hat V^{(q)}$ is a reshuffling of all terms without a derivative on $\delta(B\cdot N)$ whereas all terms with a derivative on the delta-function in \cite{Berkovits:2004px} are now moved to the $Q$-exact piece. 
The operators that enter the definition of $\hat V^{(0)}$ are of the symbolic form\footnote{
We restrict ourselves to the superparticle case and do not specify the constant coefficients, which can
 be computed as explained in the previous footnote. For more details we refer the reader to section~4 of \cite{Berkovits:2004px}.}:
\begin{eqnarray}
\label{OpI}T&=& PP \nn\\
\label{OpII}G^{A}&=& dP \nn\\
\label{OpIII}H^{AB}&=& dd + NP + JP \nn\\
\label{OpIV}K^{ABC}&=&Nd+Jd \nn\\
\label{OpV}L^{ABCD}&=& NN+JN+JJ\ .
\end{eqnarray}
The operators $T$ and $G$, which enforce the Virasoro and $\kappa$-symmetry constraints respectively, are already present without the introduction of pure spinors. On the other hand, the operators $H$, $K$ and $L$ are new since they contain $\lambda^{A}$ and $w_{A}$.\footnote{We would like to elaborate on the relation between the structure of the BRST operator and antighost field in the pure spinor approach and the NSR superstring: in the former, $Q$ contains only a single term at zero picture while the antighost $b_{B}$ is a complicated polynomial. In the latter, the BRST charge $Q_{NSR}$ can be decomposed into three pieces, whereas the antighost field $b_{NSR}$ contains only one term. However, there exists a similarity transformation which maps $Q_{NSR}$ into a single nilpotent piece and at the same time maps the field $b_{NSR}$ into a complicated expression that contains exactly the supersymmetry generator and the Virasoro constraints.} 

Using the above manipulations, we can rewrite the one-loop amplitude prescription in the following way: 
\be \label{oneH} 
{\cal A} = 
\int \frac{d\tau}{\tau} 
\Big \langle 
Z_{J} \, \prod_{P=1}^{22} Z_{B}  \, \prod_{I=1}^{23} Y_{C_{I}} 
 \hat V^{(0)}(\tau_{1}) \prod_{T=2}^{N} \int d\tau^{T} V(\tau^{T}) 
\Big\rangle \, .
\ee  
The counting of zero modes goes as in Section~\ref{section:oneL}, especially if one realizes that the expression~(\ref{newvop}) just gives a different distribution of the $\lambda$'s, $d$'s and $\theta$'s. This counting shows that only the second term in $\hat{V}^{(0)}_B$, namely $H^{AB} D_B \Phi^{(q)}_A$, contributes to the four-point
amplitude.\footnote{In the rest of this paper only the $H^{AB}$ term will be necessary; the other terms will contribute to higher point and/or higher loop amplitudes.} 

Two final remarks: {\it i)} 
Notice that all manipulations are purely algebraic and 
do not require any supplementary information such as the 
constraints coming from conformal 
invariance in string theory. In the latter case, CFT is needed 
in order to evaluate the contribution of the non-zero modes. But 
for our purposes the algebraic properties of pure spinors, together 
with Lorentz covariance, are completely sufficient. To extend the present analysis 
to the membrane one would need to know the contributions coming from the non-zero 
mode part of the theory. {\it ii)} Even if the prescription given in (\ref{oneH}) 
does not seem to be symmetric with respect to the interchange of vertex operators, 
one can check, using the descent equations for the vertex operators, that 
it is symmetric.

\section{One loop amplitudes} 

As a check on the correctness and feasibility of the prescription for loop calculations in the pure spinor formalism, in this section we compute the effective action terms induced by several one-loop amplitudes: the ten-dimensional $B\wedge X_8$, found in \cite{VafaWitten}, the eleven-dimensional $C\wedge X_8$, deduced in \cite{DLM}, and also the supersymmetric ${\cal F}^4$ in type I.

\subsection{$B\wedge X_8$ in ten dimensions}

We compute the five-point amplitude $B\wedge t_8R^4$ both in type~IIA and type~IIB superstring using the formalism of \cite{Berkovits:2004px}.

For this calculation we will only need the vertex operators
for the graviton and the $B$-field in the integrated $V^{(0,0)}$ picture,
unintegrated $U^{(1,1)}$ picture, and the mixed pictures  $V^{(1,0)}$ and
$V^{(0,1)}$. We use the results of \cite{membr,Grassi:2004ih} and refer to these
papers for complete expressions. For our purposes it is enough to take the $B$-field and Riemman curvature to be constant. The relevant vertex operators are
\begin{eqnarray}\label{vop2}
V^{(0,0)}_g &=& \int d^2z\, R_{mnpq} \, \mathcal{L}_L^{mn} \mathcal{L}_R^{pq}\, e^{ip\cdot X}\\
\nonumber
V^{(0,0)}_B &=& \int d^2z\, B_{mn} \, \partial X^m \, \bar\partial X^n
e^{ip\cdot X}\\
\nonumber
U^{(0,1)}_g &=& \int dz\, R_{mnpq}\, \mathcal{L}_L^{mn} (\lambda_R
\gamma^r\theta_R) (\theta_R \gamma_r{}^{pq}\theta_R) \, e^{ik\cdot X}\\
\nonumber
U^{(0,1)}_B &=& \int dz\, B_{mn}\, \partial X^m (\lambda_R
\gamma^n\theta_R)  \, e^{ik\cdot X}\\ 
\nonumber
U^{(1,1)}&=& (g_{mn}+ B_{mn}+\eta_{mn}\varphi) \, (\lambda_L \gamma^m\theta_L)
(\lambda_R \gamma^m\theta_R) \, e^{ip\cdot X}\ ,
\end{eqnarray}
where $\mathcal{L}_{L,R}^{mn}$ are the Lorentz generators
\begin{equation}
\mathcal{L}_L^{mn} = \frac{1}{2} (w_L \gamma^{mn}\lambda_L) + (p_L
\gamma^{mn}\theta_L) + X^{[m} \partial X^{n]}\ , 
\end{equation}
with the equivalent expression for the right-movers. Since we only need  the zero modes of fields we can, in particular, replace $\left.p_\alpha\right|_0$ by $\left.d_\alpha\right|_0 + i/2\,  \partial X_m \left.(\gamma^m \theta_L)_\alpha\right|_0$. From now on we drop the
subscript 0, although we will always mean the zero modes of the corresponding fields. So the Lorentz generator becomes 
\begin{equation}\label{LorentzL}
\hat {\mathcal{L}}_{L}^{mn} = \frac{1}{2} (w_L \gamma^{mn}\lambda_L) + (d_L
\gamma^{mn}\theta_L) + \frac{i}{2} \partial X_l (\theta_L \gamma^{lmn}\theta_L) + X^{[m} \partial X^{n]}\ . 
\end{equation}

The five-point amplitude with one $B$-field and four gravitons can be written in
three different forms depending on how the ghost-charge is distributed among the
$B$-field vertex operator and the graviton vertex operator
\begin{eqnarray}\label{FivePointI}
\mathcal{A}_1&=&  \left|\int \mu b\, Z_J\prod_{i=2}^{10}
Z_B\prod_{i=1}^{11}Y_C  \right|^2 \, U^{(1,1)}_B \,\left( V_g^{(0,0)}\right)^4\\
\mathcal{A}_2&=&  \left|\int \mu b\, Z_J\prod_{i=2}^{10}\,
Z_B\prod_{i=1}^{11}Y_C   \right|^2\,  U^{(1,1)}_g \, V_B^{(0,0)}\,\left( V_g^{(0,0)}\right)^3\\ \label{FivePointIII}
\mathcal{A}_3&=&  \left|\int \mu b\, Z_J\prod_{i=2}^{10}
Z_B\prod_{i=1}^{11}Y_C   \right|^2\, \left[ U^{(1,0)}_B \, U^{(0,1)}_g +  U^{(0,1)}_B \, U^{(1,0)}_g  \right]\left( V_g^{(0,0)}\right)^3
\end{eqnarray}
However, the result is independent on the distribution of the ghost number as can be seen in the following way: Write
$$
U^{(1,1)}= \int d^2z \, \partial\bar\partial U^{(1,1)}= \{Q_B^L, U^{(0,1)}\}= \{Q_B^R, U^{(1,0)}\}
$$
and circulate the left (right) BRST charge $Q_B^L$ ($Q_B^R$) inside the integral. Now, the picture changing operators $Z_B$ and $Z_J$ are BRST invariant but the $b$ ghost is not, its variation being $\{Q_B^L, b\}= Z_B T(z)$. Nevertheless, the latter contribution vanishes after integration over the moduli since $T$ gives a total derivative with respect to them.

Despite the equivalence of (\ref{FivePointI})-(\ref{FivePointIII}), similarly to the NSR formalism, there is a computationally preferred
distribution of ghost number, namely the one given by $\mathcal{A}_3$. Using the correlator
$$
\langle \partial X^m(z) \bar\partial X^n(0)\rangle = \pi \alpha'\, \left(\delta^{(2)}(z) -
\frac{1}{\tau_2}\right) \, \eta^{mn} \, ,$$ 
where $\tau_2 = {\rm Im} \tau$, the amplitude reduces to 
\begin{eqnarray}
\nonumber\mathcal{A}_3&=& \int \frac{d^2{\tau}}{\tau_2^{2}}\,
\left|\int [\mathcal{D}\lambda][\mathcal{D}w] \text{d}^{16}\theta \text{d}^{16}d\, \int b\,
Z_J\prod_{i=2}^{10} Z_B\prod_{i=1}^{11}Y_C \right|^2 \,\left( V_g^{(0,0)}\right)^3\\
&&\times R_{rstu}B_{[[mn}\, g_{kl]]}\,\left[ (\lambda_R \gamma^n\theta_R)
  (\theta_R \gamma^{mtu}\theta_R) 
  (\lambda_L\gamma^k\theta_L)(\theta_L \gamma^{lrs}\theta_L) \right]\,
e^{ik^{(1)}\cdot X} e^{ik^{(2)}\cdot X}
\end{eqnarray}
with $B_{[[mn}g_{kl]]}=B_{mn}g_{kl}- B_{kl}g_{mn}$. In the 0-ghost-picture vertex operator
only the part 
\begin{equation}\label{LorentzM}
\mathcal{M}_{L,R}^{mn}= \frac{1}{2} (w_{L,R} \gamma^{mn}\lambda_{L,R}) + (d_{L,R}
\gamma^{mn}\theta_{L,R}) 
\end{equation}
of the Lorentz generator $\hat{\mathcal{L}}_{L}^{mn}$ in~(\ref{LorentzL}) contributes
and  after integration over the position of the vertex operator, the leading term in $\alpha'$
expansion for $\mathcal{A}_3$ reduces to
\begin{equation}
\mathcal{A}_3=\int \frac{d^{2}\tau}{\tau_2^{2}} \, t_{L}^{kl|m_{1}\cdots m_{8}}\, 
t_{R}^{mn|n_{1}\cdots n_{8}}\ B_{[[mn}\, g_{kl]]} \, R_{m_1m_2n_1n_2} \cdots R_{m_7m_8n_7n_8} \,. \label{Ampl}
  \end{equation}
In (\ref{Ampl}) we have defined the left and right tensors\footnote{Notice that these tensors are elements of the cohomology $H^{(1)}(Q)$ as they arise from an amplitude with one vertex operator in the ghost-number-1 picture and three vertex operators in the ghost-number-0 picture.}
\begin{equation}\label{Tensort}
 t^{kl|m_{1}\cdots m_{8}}= \int [\mathcal{D}\lambda][\mathcal{D}w] \, b\, Z_J\prod_{i=2}^{10}
Z_B\prod_{i=1}^{11}Y_C  \left[(\lambda\gamma^k\theta)(\theta
\gamma^{l<m_{1}m_{2}}\theta) 
  \mathcal{M}^{m_{3}m_{4}}\cdots \mathcal{M}^{m_{7}m_{8}>}\right] ,
   \end{equation}
where $< \cdots >$ denotes symmetrization over exchanges of the pairs $(m_{2i-1}, m_{2i})$ $i=1,\dots, 4$,
   as a consequence of the many ways of choosing the distribution of the indices when performing the zero mode integration.
After integrating over the $\lambda$'s and $d$'s as in section~6.3 of
\cite{Berkovits:2004px}, one finds that this tensor reduces to 
\begin{eqnarray}\label{tL}
 t_{L}^{kl|m_{1}\cdots m_{8}}&=&\int d^{16}\theta_L 
   (\epsilon_{16}\mathcal{T}^{-1}_L)^{((\alpha_1\cdots\alpha_3))}_{[\rho_1\cdots \rho_{11}]} 
  \theta_L^{\rho_1}\cdots \theta_L^{\rho_{11}} \\
\nonumber&&\times\,
(\gamma_{p_1\cdots p_3}{}^{<m_{7}m_{8}})_{\alpha_2\alpha_3} (\gamma^k\theta_L)_{\alpha_1}(\theta_L
  \gamma^{lm_{1}m_{2}}\theta_L)(\theta_L\gamma^{m_3m_4} \gamma^{p_1\cdots p_3} \gamma^{m_5m_6>} \theta_L)\\
\nonumber&=&
(\mathcal{T}^{-1}_L)^{((\alpha_1\cdots\alpha_3))}_{[\rho_1\cdots \rho_5]}\, (\gamma_{p_1\cdots p_3}{}^{<m_{7}m_{8}})_{\alpha_2\alpha_3} (\gamma^k)_{\alpha_1}{}^{[\rho_1}( \gamma^{lm_{1}m_{2}})^{\rho_2\rho_3}
(\gamma^{m_5m_6} \gamma^{p_1\cdots p_3} \gamma^{m_7m_8>})^{\rho_4\rho_5]} \,.
\end{eqnarray}
The tensors $\mathcal{T}_{L,R}^{-1}$ are given by the tree-level normalization~(\ref{normtree}):
\begin{eqnarray}
\left(\mathcal{T}_L^{-1}\right)^{((\alpha_1\cdots
\alpha_3))}_{[\rho_1\cdots \rho_5]}&=&\langle \lambda^{\alpha_1}\lambda^{\alpha_2}\lambda^{\alpha_3} \theta_{\rho_1}\cdots \theta_{\rho_5}\rangle_{g=0}\\
\nonumber&=&\Pi^{\alpha_{1}\cdots\alpha_{3}}_{\beta_{1}\cdots \beta_{3}} 
(\gamma^{n_1})^{(\beta_1}{}_{[\rho_1}(\gamma^{n_2})^{\beta_2}{}_{\rho_2}(\gamma^{n_3})^{\beta_3)}{}_{\rho_3}(\gamma^{n_1\cdots
n_3})_{\rho_4\rho_5]} \, ,
\end{eqnarray}
where 
\begin{equation}\label{PiI}
\Pi^{((\alpha_{1}\cdots\alpha_{3}))}_{\beta_{1}\cdots \beta_{3}} =
 \delta^{(\alpha_{1}}_{\beta_{1}} \cdots\delta^{\alpha_{3})}_{\beta_{3}}  - 
 \frac{1}{6} \, (\gamma^{m})^{(\alpha_{1}\alpha_{2}} \delta^{\alpha_{3})}_{(\beta_{1}} \, (\gamma_{m})_{\beta_{2}\beta_{3})}
\end{equation}
projects on $\gamma_{m}$-traceless symmetric tensors, i.e. $(\gamma_{n})_{\alpha_{1}\alpha_{2}}\, \Pi^{\alpha_{1}\cdots\alpha_{3}}_{\beta_{1}\cdots \beta_{3}} =0$. 

Note that the tensor $t_{L}$, defined as the one-loop saturation of zero modes in~(\ref{tL}), can also be written as a tree-level
correlator in the following way:
\begin{equation}
 t_{L}^{kl|m_{1}\cdots m_{8}}= \langle  (\lambda\gamma^k\theta)
 (\theta\gamma^{l<m_{1}m_{2}}\theta) (\lambda\gamma_{p_1\cdots p_3}{}^{m_{3}m_{4}}\lambda)
(\theta\gamma^{m_5m_6} \gamma^{p_1\cdots p_3} \gamma^{m_7m_8>}\theta)\rangle_{g=0} \, .
\end{equation}
We only need the contractions of the $k l$ indices of this tensor with the metric and with the $B$-field. To bring the expression for $t_{L}$ in a manageable form, we utilize the Fierz identity
\begin{equation}\label{FI}
-\frac{1}{3!\, 16}\, (\lambda
\gamma_{p_{1}\cdots p_{3}}{}^{m_{1}m_{2}}\lambda)(\theta\gamma_{m_{3}m_{4}}\gamma^{p_{1}\cdots
p_{3}}\gamma_{m_{5}m_{6}}\theta)=(\lambda
\gamma^{[m_{1}}\gamma_{m_{3}m_{4}}\theta)
(\lambda\gamma^{m_{2}]}\gamma_{m_{5}m_{6}}\theta) \, .
\end{equation}
Then we use the following identities:
\begin{eqnarray}
\langle(\lambda\gamma^{m_{1}\cdots m_{3}}\theta) (\lambda\gamma^{m_{4}\cdots
m_{6}}\theta)(\lambda\gamma^{m_{7}}\theta)
 (\theta\gamma_{m_{8}\cdots m_{10}}\theta) \rangle_{g=0}&=&
\pm\frac{ 1}{3!}\,\epsilon_{10}^{m_{1}\cdots m_{10}} \, , \label{LI}\\
\langle(\lambda\gamma^{m_{1}}\theta)\cdots (\lambda\gamma^{m_{3}}\theta)
 (\theta\gamma_{n_{1}\cdots n_{3}}\theta) \rangle_{g=0}&=& \frac{1}{120}\,\delta^{m_{1}\cdots
m_{3}}_{n_{1}\cdots n_{3}} \, , \nn\\
 \langle(\lambda\gamma^{m}\theta) (\lambda\gamma^{n}\theta)
 (\lambda\gamma^{p_{1}\cdots p_{3}}\theta)
 (\theta\gamma_{q_{1}\cdots q_{3}}\theta) \rangle_{g=0}&=&
\frac{1}{30} \left(\delta^{mp_{1}\cdots p_{3}}_{nq_{1}\cdots
q_{3}}-\delta^{np_{1}\cdots p_{3}}_{mq_{1}\cdots q_{3}}-\frac{1}{4}
\delta^{m}_{n}\, \delta^{p_{1}\cdots p_{3}}_{q_{1}\cdots q_{3}}\right)\, . \nn
\eea
The sign in the first equation is determined by the chirality of the ten-dimensional
complex Weyl fermions  $\theta$.
With the help of (\ref{FI})-(\ref{LI}) and~(\ref{teightI}), it
is not difficult to check that
 \begin{eqnarray}
g_{kl}\, t_{L,R}^{kl|m_{1}\cdots m_{8}}&=&t_{8}^{m_{1}\cdots m_{8}}\\
 B_{kl}\, t_{L,R}^{kl|m_{1}\cdots m_{8}}&=&B_{kl}\, (t_{10}^{klm_{1}\cdots m_{8}}\pm\epsilon_{10}^{klm_{1}\cdots m_{8}}) \, .
 \end{eqnarray}
The tensor $t_{10}$ is only composed of $\eta_{MN}$'s and contracting it with a set of generic two-forms $F^i_{mn}$, $i=1,...,4$, gives
$$
t_{10}^{klm_{1}\cdots m_{8}}B_{kl} F_{m_{1}m_{2}}^{1}\cdots F^{4}_{m_{7}m_{8}}=a_{1}\text{tr}(B F^{1}) \text{tr}(F^{2}F^{3}F^{4})+a_{2} \text{tr}(BF^{1}\cdots F^{4}) + perms \, ,
$$
symmetrized in the exchange of the $F^{i}$. However, the antisymmetry of $F^{i}$ and $B$ makes this expression to vanish.
The sign in front of the $\epsilon_{10}$-term is related to the chirality of the $\theta$ variables. Since for 
type IIA/IIB $\theta_{L}$ and $\theta_{R}$ have the opposite/same chiralities, by antisymmetry of the $B$-field the amplitude~(\ref{FivePointIII}) vanishes for type IIB and gives the $B\wedge t_{8}t_8R^{4}$ term for type IIA.

\subsection{$C_3\wedge X_8$ in eleven dimensions}
The zero mode counting of Section~\ref{section:oneL} shows that the first non
zero amplitudes are with four external states. In particular, the one-loop
four-graviton scattering, that produces the $R^4$ term, is non
vanishing.\footnote{This and related amplitudes will be discussed elsewhere \cite{AGV}.}
In this subsection we will be interested in the supersymmetric partner of the
latter, namely the five-point amplitude with four gravitons and a $C_{3}$-field.

 The relevant vertex operators are restrictions of the vertex operators given in Subsection~\ref{section:vop} to constant curvature $R_{MNPQ}$. The graviton vertex operators are given by
\begin{eqnarray}
\nonumber V_{R}&=& \int d\tau\, R_{MNPQ}\, \hat{\mathcal{L}}^{MN}
\hat{\mathcal{L}}^{PQ}\, e^{ik\cdot X}\\
U^{(1)}_{R}&=& R_{MNPQ}\, \hat{\mathcal{L}}^{MN} (\lambda \Gamma^R\theta) (\theta
\Gamma_R{}^{PQ}\theta)\, e^{ik\cdot X}\ ,
\end{eqnarray}
where $\hat{\mathcal{L}}^{MN}$ stands again for the zero mode restriction of the Lorentz generators
\begin{equation}
\hat{\mathcal{L}}^{MN} = N^{MN} + d\Gamma^{MN} \theta + \frac{i}{2}P_{L}(\theta\Gamma^{LMN}\theta)+ P^{[M} x^{N]} \, .
\end{equation}
Unfortunatly the vertex operator for the $C_3$-field is not so simple because the superparticle is not charged under it. Looking at the vertex operators of section~\ref{section:vop}, and in particular at the expressions for the super-vielbein components in (\ref{superVielbein}), it appears that the field strength $G_4$ comes always in combination with the connection $\omega_M{}^{rs}$ as 
\begin{equation}
V_{G_{4}+\omega}= \int d\tau \,  \left[ \omega_{M}{}^{rs}(d\Gamma_{rs}\theta)+(d\mathcal{T}_{M}\cdot G_{4}\theta)\right]\,
P^{M}\, e^{ik\cdot X}\ .
\end{equation}
This means that the 5 points amplitude will read 
$$
A_5 \sim (G_4 + \omega)^2 \,  R^3 \sim C_3 R^4 + R^4 + (G_{4})^2 \, R^{3} \, ,
$$
which shows that supersymmetry relates automatically the $R^4$ term with the
Chern-Simons $C_3\wedge t_8 R^4$ one.\footnote{Note that $R^4$  contains two
contributions: one is the non-linear completion of the four-point $t_8t_8 R^4$
term and the other gives $\epsilon_{11}\epsilon_{11}R^4$.}

As an illustration we consider the amplitude composed of three integrated vertex
operators for the graviton, one for the spin-connection, one for $G_{4}$ and one
unintegrated vertex operator for the graviton. It is given by
\begin{eqnarray}\label{Xeight}
\mathcal{A} &=& 
R_{M_1M_2N_1N_2}\cdots R_{M_{5}M_{6}N_{5}N_{6}} \,
\omega_{M_{7}}{}^{N_{7}N_{8}}\, G_{R_{1}\cdots R_{4}}\\
\nonumber &&\int [\mathcal{D}\lambda][\mathcal{D}w] \text{d}^{32}\theta \text{d}^{32}d\,
\int \frac{d\tau}{\tau}
b\,  \prod_{i=1}^{4} d\tau_{i}\,  Z_{J}
\prod_{I=2}^{22} Z_{B}\prod_{I=1}^{32} Y_{C}\\
\nonumber && {\mathcal{M}}^{M_{1}M_{2}}\cdots{\mathcal{M}}^{M_{5}M_{6}}
{\mathcal{M}}^{N_{1}N_{2}}{\mathcal{M}}^{N_{3}N_{4}}\\
 \nonumber&&  (\lambda \Gamma^{R}\theta)(\theta\Gamma^{RN_{5}N_{6}}\theta)\, 
 P^{M_{7}}N^{N_7N_8}\, (d\mathcal{T}_{S}{}^{R_{1}\cdots R_{4}}\theta)P^{S}\,
\prod_{1\leq i<j \leq 5} e^{ik^{(i)}\cdot k^{(j)}\, G_{ij}}
\, ,
\end{eqnarray}
where, as for the superstring, only the part 
$$
\mathcal{M}^{MN}= N^{MN} + (d\Gamma^{MN}\theta)
$$
of the Lorentz generator $\hat{\mathcal{L}}^{MN}$ contributes from the integrated vertex operators.
The $b$-ghost has to soak up the zero modes of one Lorentz generator \cite{Berkovits:2004px},
which we choose to take from the vertex operator for the spin connection. Using that for the on-shell superparticule $P^{M}=\dot X^{M}$ and also the world-line expression for the one-loop propagator \cite{Strassler:1992zr}
$$
\langle X^{M}(\tau_{i}) X^{N}(\tau_{j}) \rangle_{g=1}= \eta^{MN}\, G_{ij}=
\frac{\eta^{MN}}{2}\left(|\tau_{i}-\tau_{j}|- (\tau_{i}-\tau_{j})^{2}
+ \text{const.}\right)
$$
and integrating over the position of the vertex operators, we obtain the following result for the
amplitude at leading order in the momenta
\begin{equation}
\mathcal{A} = \int \frac{d\tau}{\tau^{3/2}}\, t_{11}^{SM_{1}\cdots M_{6}R_{1}\cdots R_{4}N_{1}\cdots N_{8}} 
R_{M_1M_2N_1N_2}\cdots R_{M_{5}M_{6}N_{5}N_{6}} \,
\omega_{S}{}^{N_{7}N_{8}}\, G_{R_{1}\cdots R_{4}}
\end{equation}
with
\begin{eqnarray}
&&t_{11}^{SM_{1}\cdots M_{6}R_{1}\cdots R_{4}N_{1}\cdots N_{8}} =\int
[\mathcal{D}\lambda][\mathcal{D}w]\text{d}^{32}\theta \text{d}^{32}d\,
\int  b\, Z_{J} \prod_{I=2}^{22} Z_{B}\prod_{I=1}^{32} Y_{C}\\
\nonumber&&\times \left[N^{N_{7}N_{8}}
{\mathcal{M}}^{M_{1}M_{2}}\cdots{\mathcal{M}}^{M_{5}M_{6}}
{\mathcal{M}}^{N_{1}N_{2}}{\mathcal{M}}^{N_{3}N_{4}}\,(\lambda
\Gamma^{R}\theta)(\theta\Gamma^{RN_{5}N_{6}}\theta)\, 
 \, (d\mathcal{T}_{M_{7}}{}^{R_{1}\cdots R_{4}}\theta)+sym.\right] \, ,
\end{eqnarray}
where symmetrization between the $(M_{2i-1},M_{2i})$ and $(N_{2i-1},N_{2i})$ pairs is understood.
The $b$-ghost insertion requires one $N^{MN}$ zero mode from the Lorentz generator $\mathcal{M}^{MN}$
and we have to get 5 $d$'s and 9 $\theta$'s to saturate the fermionic zero modes. Similarly to the superstring, with the help of (\ref{inverseT}) we can rewrite this tensor as a tree-level correlator
\begin{eqnarray}
&&t_{11}^{SM_{1}\cdots M_{6}R_{1}\cdots R_{4}N_{1}\cdots N_{8}}=
\langle (\lambda\Gamma^{R} \theta)(\theta \Gamma^{RN_{1}N_{2}}\theta)\\
\nonumber&\times&(\lambda \Gamma^{N_{3}N_{4}} \Gamma_{P_{1}\cdots P_{5}}\Gamma^{N_{5}N_{6}}\lambda) 
(\lambda \Gamma_{P_{6}P_{7}}\lambda )(\lambda\Gamma_{P_{8}P_{9}}\lambda )\\
\nonumber&\times&(\theta\mathcal{T}_{M_{7}}{}^{R_{1}\cdots
R_{4}}\Gamma^{P_{1}\cdots P_{3}}\Gamma^{N_{7}N_{8}}\theta) 
(\theta \Gamma^{M_{1}M_{2}}\Gamma^{P_{4}\cdots P_{6}}\Gamma^{M_{3}M_{4}}\theta )
(\theta \Gamma^{M_{5}M_{6}}\Gamma^{P_{7}\cdots P_{9}}\Gamma^{M_{7}M_{8}}\theta )\rangle_{g=0} \, ,
\end{eqnarray}
symmetrized over all exchanges of the pairs $(M_{2i-1},M_{2i})$ and $(N_{2i-1},N_{2i})$ for $i=1,\dots,
4$. By Fierzing and using the tree-level normalization~(\ref{normtree}), one can see that the last expression has an expansion of the following general 
form: 
\begin{equation}
\sum_{n_{1}+ n_{2}+ \cdots =8\atop n_{i}\geq 0}\, c_{\{n_{i}\}}\,
\text{tr}\left((\Gamma_{[2]})^{n_{1}}\right)\cdots
\text{tr}\left((\Gamma_{[2]})^{n_{7}}\right)
\text{tr}\left((\Gamma_{[2]})^{n_{8}}\mathcal{T}_{S}{}^{R_{1}\cdots
R_{4}}\right)\, ,
\end{equation}
where $\Gamma_{[2]}$ represents any of $\Gamma^{[M_{2i-1}M_{2i}]}$ or
$\Gamma^{[N_{2i-1}N_{2i}]}$. The last multiplier in this expression is nonzero
only for $n_{8}=4$, picking the
 $\eta_{M_{7}}^{[R_{1}}\Gamma^{R_{2}\cdots R_{4}]}$
from
 $\mathcal{T}_{M_{7}}{}^{R_{1}\cdots R_{4}}$, to give an eleven dimensional $\epsilon$-tensor
$\epsilon_{11}^{R_{2}\cdots R_{4}\cdots }$. Integrating by parts the derivative on the
four-form and relabelling the pairs of $M$ and $N$ indices, 
 we obtain for the amplitude
 \begin{equation}
\mathcal{A}= \epsilon_{11}^{M_{1}\cdots M_{11}}
C_{M_{9}\cdots
M_{11}} R_{M_{1}M_{2}N_{1}N_{2}}\cdots R_{M_{7}M_{8}N_{7}N_{8}} 
 \left(
\text{tr}\left((\Gamma^{NN})^{4}\right)-\frac{1}{16}
\left(\text{tr}(\Gamma^{NN})^{2}\right)^{2}\right)\, .
 \end{equation}
 Finally using~(\ref{teightII}) with $\Delta=32$, we find that 
 this amplitude gives $C_{3}\wedge
t_{8}R^4$.

Obtaining the correct structure is very satisfying, but we should note that we haven't addressed here the issue of the UV regularization of the 11d answer. In the light-cone computation of \cite{GGV} it was resolved by considering M-theory on $R^{(8,1)} \times T^2$ and requiring that the reductions to IIA and IIB give T-dual answers. In the present pure spinor  formulation such a program is a nontrivial task and we intend to come back to it in the future. 

\subsection{Open string amplitudes in ten dimensions} 

In the previous subsections we verified the correctness of the definition of the one-loop measure by recovering the known $B\wedge R^4$ and $C_{3}\wedge R^4$ terms in closed srting theory and eleven-dimensional supergravity respectively. In the current subsection we provide yet another check on the measure by comparing with known results for a simpler case, namely one loop in 10d open string theory.
More precisely, we compute the supersymmetric ${\cal F}^{4}$ contribution to the effective action. 
By that we mean all three terms: four gluons; two gluons and 
two gluinos; four gluinos. 

Again we restrict ourselves to the massless states. The general expression for the covariant supervertex operator of 10d Super Yang-Mills is\footnote{For simplicity we take the lowest component of the superfield $F_{mn}$, which is the gauge field strength, to be constant.} 
\be \label{VOp} 
{\mathcal V}^{(0)} = \dot{\theta}^{\a} A_{\a}(x,\theta) + \Pi^m A_m(x,\theta) +
d_{\a} W^{\a} (x, \theta) + (w \g^{mn} \l)F_{mn} (x, \theta) \, . 
\ee 
Using the field equations and Bianchi identities, one can derive the relations \cite{Witten:1985nt, HS}:
\bea \label{rec}
&&D_{( \a} A_{\b )} = \g^m_{\a \b} A_m \nn \\ 
&&D_{\a} A_m - \partial_m A_{\a} = \g_{m \a \b} W^{\b} \nn \\ 
&&D_{\a} W^{\b} = \frac{1}{4} \g^{mn}{}_{\a}{}^{\b} F_{mn} \nn \\ 
&&D_{\a} F_{mn} = (\g_{[m} \partial_{n]} W)_{\a} \, .
\eea 
Together with the gauge choice 
\be 
\theta^{\a} A_{\a} = 0 \, ,
\ee 
equations (\ref{rec}) allow one to obtain the full expansion in powers of $\theta$ of the superfields $A_{\a}$, $A_m$, $W^{\a}$ and $F_{mn}$ following the iterative procedure of \cite{HS, ORRT}. The only input is the identification of the lowest components as 
\be 
A_m|_{\theta = 0} = a_m(x) \, , \qquad W^{\a}|_{\theta = 0} = u^{\a} (x) \, , \qquad F_{mn}|_{\theta = 0} = f_{mn} \, , 
\ee 
where $a_m$ is the gauge field and $u^{\a}$ is the gluino. Using 
that for a constant field strength $a_m = f_{mn} x^n$, 
one finds for the first few terms of these expansions: 
\bea 
A_{\a} &=& a_m (\g^m \theta)_{\b} + (\g^m \theta) (u \g_m \theta) + ... \nn \\ 
A_m &=& a_m + u \g^m \theta + f_{pq} \theta \g^{mpq} \theta + ... \nn \\ 
W^{\a} &=& u^{\a} + \frac{1}{4} f_{mn} (\g^{mn} \theta)^{\a} + \cdots \nn \\ 
F_{mn} &=& f_{mn} - \theta \g_{[m} \partial_{n]} u + \cdots 
\eea 
Substituting the last equations in (\ref{VOp}), we obtain for the supervertex operator (on shell i.e. using also $\dot{\theta} = 0$ and $\dot{x}^m = P^m$): 
\be \label{VerOp} 
{\mathcal V}^{(0)} = u^{\a} q_{\a} + f_{mn} (\mathcal{M}^{mn} + P^{[m} x^{n]}) - (\theta \g_{[m} \partial_{n]} u) (\frac{1}{2} \mathcal{M}^{mn} + \frac{1}{3}P_p \theta \g^{pmn} \theta) \, , 
\ee 
where $\mathcal{M}^{mn}$ is defined as in~(\ref{LorentzM}) and $q_{\a}$ is the supersymmetry generator. 

Now we will consider the four-point amplitude at one loop. Our goal will be to check if the entire effective SYM action to order $\a^{\prime 2}$ \cite{CNT} is reproduced from the computational rules of the pure spinor formalism. There are three cases for the external particles: four gluons; two gluons and two gluinos; four gluinos. We will handle all of them in turn.

For the case of four external gauge fields we take the following vertex operators: one $U^{(1)}_f = f_{pq} \l \g_{r} \t \t\g^{pqr}\t$ and three ${\cal V}^{(0)}_f = f_{pq} (\mathcal{M}^{pq} + P^{[ p} x^{q ]})$. Using the results of the previous sections, the one-loop prescription gives at leading order in the $\alpha'$ expansion 
\begin{eqnarray}\label{eeP}
&&\langle Z_{J} \prod_{1}^{10}Z_{B} \prod_{1}^{11} Y_{C}(\l,\t) \times 
\Big[  G^{\a} + H^{\a\b} D_{\b} + K^{\a\b\g} D_{\b} D_{\g} + 
L^{\a\b\g\d} D_{\b} D_{\g} D_{\d} \Big] \frac{\partial}{\partial \l^{\a}}\nonumber \\
&& U^{(1)}_f(\l,\t)
 (\int d\tau {\mathcal V}^{(0)}_f)^{3} \rangle =  t_{8} f^{4}+\mathcal{O}({\alpha'}^2) \, ,
\end{eqnarray}
which coincides exactly with the $f^4$ term in (3.7) of \cite{CNT}.

To describe the interaction of two gauge fields and two gluinos, in addition to a $U^{(1)}_f$ and a ${\cal V}^{(0)}_f$ we also need two gluino vertex operators ${\cal V}^{(0)}_u = u^{\a} q_{\a} - (\theta \g_{[m} \partial_{n]} u) (\frac{1}{2} \mathcal{M}^{mn} + \frac{1}{3}P_p \theta \g^{pmn} \theta)$. By counting the zero modes of the various fields, it is easy to convince oneself that the only nonzero contribution comes from
\be \label{Lag1}
f_{pq} u^{\a} (\g_m \partial_n u)_{\b} \, \langle \langle U^{(1)}_f
\mathcal{M}^{pq} \,\theta^{\b} \frac{1}{2} \mathcal{M}^{mn} q_{\a} \rangle \rangle_{g=1} \, , 
\ee 
where actually only the $p \,\theta$ term of ${\cal M}$ contributes. In (\ref{Lag1}) we have abbreviated by $\langle \langle ... \rangle \rangle$ the full one-loop measure, including the picture changing operators and the $b$-ghost. This last expression gives the effective interaction
\be 
{\cal L}_{f^2 u^2} = - 2 [ f_{ij} f_{pq} u^{\a} (\g_m \partial_n u)_{\b} ] \, \d^{m [ i|} \d^{n [p} \d^{q] |j]} \d_{\a}{}^{\b} \, , 
\ee
which matches precisely the corresponding term in (3.7) of \cite{CNT}.

Finally, we turn to the four-gluino one-loop amplitude. Now we need three ${\cal V}^{(0)}_u$'s and one unintegrated vertex operator for $u^{\a}$: $U^{(1)}_u = \l \g^m \theta \, u \g_m \theta$. Following the same steps as before we find for the amplitude
\bea 
{\cal L}_{u^4} &=& u^{\b} \partial_n u^{\g} \partial_q u^{\d} \,\, \langle \langle \,(\g_p \theta)_{\d} \frac{1}{2} \mathcal{M}^{pq} (\g_m \theta)_{\g} \frac{1}{2} \mathcal{M}^{mn} U^{(1)}_u q_{\b} \,\rangle\rangle_{g=1} \nn\\
&=& u^{\a} u^{\b} \partial_n u^{\g} \partial_q u^{\d} \left( \frac{1}{180} \g_{\a \b}^{abc} \d^{n q} \g_{\g \d,\, abc} + \frac{3}{10} \g_{\a \b}^{abc} \d^{n}{}_{[ a} \g_{\g \d , b} \d^q{}_{c ]} \right) \, , 
\eea 
which again agrees with the corresponding term in (3.7) of \cite{CNT}.

\section{Equivalence between spinorial and BRST cohomology}

\subsection{Deformations and cohomology}

So far we have been reconstructing the effective action from scattering amplitudes. However, there is another way of achieving the same goal. Namely, one can study the possible deformations of the equations of motion, 
consistent with the symmetries of the theory, and then integrate them to terms in the effective action. This is essentially the spinorial cohomology approach of \cite{swedes} $-$ \cite{HoweTsimpis}, pioneered back in \cite{BLBPT}. 

For instance, in the case of SYM theory the presence of radiative corrections is equivalent to deforming the superspace constraints as
\cite{Witten:1985nt,Nilsson:1985cm,Atick:1985iy,Nilsson:1986cz,Gates:1985wh,Gates:1986is,Howe:1991mf,Howe:1991bx,CNT}:
\begin{equation}\label{defA}
F_{\a\b} = D_{(\a} A_{\b)} - \g^{m}_{\a\b} A_{m} = J_{(\a\b)} \,.
\end{equation}
The right hand side contains the symmetric gamma-traceless
spinorial 2-form $J_{(\a\b)}(W,F)$ which \ is a function of a gauge invariant combination of the spinorial field
strength $W^{\a}$ and the curvature $F_{mn}$. In order that
$J_{(\a\b)}$ is a consistent deformation it must be closed with respect
to the spinorial differential $D_{\a}$:
\be\label{defB}
D_{(\a} J_{\beta\gamma)} = \g^{m}_{(\a \beta} J_{\gamma) m} \, ,
\ee
where $J_{\gamma m}$ takes into account that the left
hand side is projected on the gamma-traceless part. An efficient
way to rewrite the above equation is to use the pure spinor
BRST language. So we replace (\ref{defA}) and (\ref{defB}) with
\be\label{defC}
\{Q, U^{(1)} \} = U^{(2)}\,, ~~~~~~ \{Q, U^{(2)}\} =0\,,
\ee
where $U^{(1)} = \l^{\a}A_{\a}$ and $U^{(2)} = \l^{\a}\l^{\b}
J_{\a\b}[U^{(1)}]$.
The functional dependence of $U^{(2)}$ reminds that the components
of this superfield depend upon the gauge invariant
combinations of the superfields in $U^{(1)}$. Notice that
$J_{(\a\b)}$ automatically contains only the 5-form part; the vector part is absent because of the pure spinor constraints \cite{Howe:1991mf,Howe:1991bx,CNT}. Note also that if
$U^{(2)}$ can be expressed as a $Q$-exact variation of some
superfield $\Omega[U^{(1)}]$, then it can be reabsorbed into
a complicated non-linear redefinition of the superfield $A_{\a}$. So, then the nontrivial $U^{(2)}$'s are given by the cohomology classes of $H^{(2)}(Q)$.

In the same way, we can consider the case of
11d supergravity. The (linearized) equations of motion can be
written in terms of a gamma-traceless
spinorial superfield $A_{ABC}$ as $\Gamma^{AB}_{MNPQR} D_A A_{BCD} = 0$ and $\Gamma^{AB}_{MN} D_A A_{BCD} = 0$. The superfield
$A_{ABC}$ defines a vertex operator 
$U^{(3)} = \l^{A} \l^{B} \l^{C} A_{ABC}$ from which it is
easy to see the invariance under
$\delta A_{ABC} = \G^{M}_{(AB} \Omega_{C)M}$.
Other possible deformations of the linearized equations of motion
can be parametrized by a vertex operator of the form
$U^{(4)} = \l^{A} \l^{B} \l^{C} \l^{D} G_{ABCD}[U^{(3)}]$
where the superfield $G_{ABCD}$ is the symmetric gamma-traceless
part of the 4-superform $G$. Again the functional dependence
of $U^{(4)}$ is through gauge invariant combinations of the
fields appearing in the vertex $U^{(3)}$.
Then, one has
\be\label{defE}
\{Q, U^{(3)} \} = U^{(4)}\,, ~~~~~  \{Q, U^{(4)}\} = 0\,.
\ee
This implies that $U^{(4)} \in H^{(4)}(Q)$. This group corresponds to the spinorial cohomology group
$H^{(0,4)}_{\tau}$ studied in \cite{swedes,CNT11d,HoweTsimpis,TsimpisDeform}. In the Berkovits'
pure spinor language 
the components of $U^{(4)}$ are made out of the antifields of the supergravity fields~\cite{membr}. 
Here the elements of the cohomology 
$H^{(4)}(Q)$ are viewed as functions of $U^{(3)}$. 
 The linearized
 equations of motion
for the antifields, obtained from the condition $Q U^{(4)}=0$, are \cite{membr}:
$$
\partial^{M} g^{*}_{MN} - \frac{1}{2} \partial_{N} (\eta^{MP} g^{MP})=0 \, , \qquad \partial^{M} C^{*}_{MNP}=0 \, .
$$
 At the linearized level one
can identify
 fields with their antifields: $g^{*}_{MN}=g_{MN}$ , $C^{*}_{MNP}= C_{MNP}$.
 This has the effect of fixing the gauge symmetries of the physical fields since the equations of motion
 of the antifields are the gauge fixing conditions of the fields. The above identification fixes the harmonic gauge for the graviton
and the $\partial^{M} C_{MNP}=0$ gauge for the three-form. In these gauges the equations 
of motion become $\partial^{2} g_{MN}= \partial^{2} C_{MNP}=0$. It is interesting to note that in this language the $\mathcal{O}(\ell_{P}^{3})$ topological deformation of the equations of motion studied in \cite{TsimpisDeform} corresponds to a non-linear 
 correction to the identification between fields and antifields, namely
 \begin{eqnarray}\label{identification}
C^{*}_{MNP} &=& C_{MNP} +\beta \, (\ell_{P})^3 \, (\Omega_{L})_{MNP} \qquad {\rm with} \qquad d\Omega_{L} = \text{tr}(R\wedge R) \,,\\
\nonumber G^{*}_{MNPQ}&=&G_{MNPQ}+ \beta \,(\ell_{P})^3\, \text{tr}(R\wedge R) \
.
 \end{eqnarray}

At this point we can see the limitations of the superparticle approach as a zero mode approximation to the supermembrane. Namely, since the vertex operator $U^{(4)}$ never enters in the definition of tree-level and loop superparticle amplitudes, no information about the value of the parameter $\beta$ can be extracted from the present formalism. It is necessary to do a membrane amplitude computation or use topological arguments as in~\cite{Witten:1996md} to fix $\beta$.


\subsection{Spinorial cohomology from an extended BRST operator}\label{section:coho}
                                                                        
The spinorial cohomology group $H^{(0,4)}_{F}$ of \cite{HoweTsimpis} is
defined by the action of a fermionic derivative $d_{F}[\omega]=[d_{1}\omega]$ on the elements
of $H^{(0,4)}_{\tau}$ associated with the $\tau_0$ part of the full differential $d =
d_{0} + d_{1} + \tau_{0} + \tau_{1}$. Here we recover the same cohomology by
constructing an extended BRST operator out of the full $d$. We view this
procedure as more complete than the somewhat artificial truncation to $\tau_0$,
although as we show below the two definitions are ultimately equivalent.

To achieve the above goal, we introduce a new type of anticommuting vector ghost
$\xi^{M}$ with the help of which we can convert the differential $d$ into a BRST
operator. In curved space the four parts of $d$ act as:
  \def\o{\omega}\def\c{\omega}
  \def\nab{\nabla}
  \bea
  (d_0 \o)_{M_1\ldots M_{p+1} A_1\ldots A_q}&=&
  \nab_{[M_1} \o_{M_2\ldots M_{p+1}] A_1 \ldots A_q}
+\frac{p}{2}T_{[M_1 M_2}{}^R \o_{|R|M_3\ldots M_{p+1}]
A_1\ldots A_q} \nn\\
&\phantom{=}&+ q(-1)^pT_{[M_1(A_1}{}^{C}\o_{M_2\ldots
M_{p+1}]|C|A_2\ldots A_q)}\\
&&\nn\\
(d_1 \o)_{M_1\ldots M_{p}A_1\ldots A_{q+1}}&=&
  (-1)^p\nab_{(A_1} \o_{M_1\ldots M_{p}A_2 \ldots A_{q+1})}
+\frac{q}{2}T_{(A_1 A_2}{}^{C} \o_{M_1\ldots M_{p}]|C|A_3\ldots
A_{q+1})} \nn\\
&\phantom{=}&+ p(-1)^p  T_{(A_1[M_1}{}^{R}\o_{|R|M_2\ldots
M_{p}]A_2\ldots A_{q+1})}\\
&&\nn\\
(\tau_0\o)_{M_1\ldots M_{p-1}A_1\ldots A_{q+2}}&=&
\frac{p}{2}T_{(A_1 A_2}{}^R\o_{|R|M_1\ldots M_{p-1}A_3\ldots
A_{q+2})}\\
&&\nn\\
(\tau_1\o)_{M_1\ldots M_{p+2}A_1\ldots A_{q-1}}&=&
\frac{q}{2}T_{[M_1 M_2}{}^{B}\o_{M_3\ldots A_{p+2}]|B|A_1\ldots
B_{q-1})} \,\, ,
\eea
where $\nabla$ is the covariant derivative, $\omega$ is a supeform belonging to $\Omega^{(p,q)}$ and $T_{\Omega \Sigma}^{~~\Theta}$ is the supertorsion. We consider flat space and therefore the only non-vanishing components of the latter are $T_{AB}^{~~M} = - i \g_{AB}^{~~M}$.
                              
Contracting the different pieces of the
differential operator $d$ with the ghost fields $\l^{A}$ and
$\xi^{M}$ (and the conjugate momenta $w_{A}, \beta_{M}$)
one finally gets the BRST operator
\bea\label{BRSTone}
Q &=&   \sum_{n=-1}^{2} Q_{n} =
\l^{A} \l^{B} T_{AB}^{~~R} \beta_{R} +
\left(\l^{A} \l^{B} T_{AB}^{~~C} w_{C} +
\l^{A} \xi^{N} T_{AN}^{~~R} \beta_{R} + \l^{A} \nabla_{A}\right) +
\nn\\
&&~~~~~~~~~
+\left(
\xi^{M} \l^{A} T_{MA}^{~~B} w_{B} +
\xi^{M} \xi^{N} T_{MN}^{~~R} \beta_{R} + \xi^{M} \nabla_{M}\right) +
\xi^{M} \xi^{N} T_{MN}^{~~A} w_{A} \, ,
\eea
where the index $n$ of the different pieces of $Q$ denotes
the grading obtained by summing the gradings of the
corresponding ghost fields. The latter have gradings zero
for $\l^{A}, w_{A}$ and $(1,-1)$ for $\xi^{M}, \beta_{M}$. In the
present context the grading coincides with the engineering dimension
of the variables $P_{M}$ and $d_{A}$. Due to the properties
of the covariant derivatives and by the defintion of the
supertorsion $T$, one can check that $Q$ is nilpotent.
                                                                                
The forms of type $(p,q)$ are mapped into vertex operators
$U^{(p,q)}$
with $p$ $\l^{A}$-ghosts and $q$ $\xi^{M}$-ghosts and the action of the BRST operator is
\be\label{cicA}
Q: U^{(p,q)} \rightarrow U^{(p+2,q-1)}
\oplus U^{(p+1,q)} \oplus U^{(p,q+1)} \oplus U^{(p-1,q+2)} \, .
\ee
The total
grading is given by $q$. Notice that the space has a double filtration:
the grading number (which is given by the number $\xi^{M}$ ghosts) and the total ghost number (form number)
which is given by the sum $p+q$. The BRST operator $Q$ raises
by one unit the ghost number. It is easy to check that, due to
the Poincar\'e lemma, all the cohomology groups $H^{(p,q)}(Q)$ are
empty, except $H^{(0,0)}(Q)$ which contains only the constant
functions. So any $Q$-closed form is also $Q$-exact. 
                                             
On the other hand, the double filtration means that it is possible
to introduce a new BRST operator which has a definite grading number,
is nilpotent and has vanishing cohomology. For that purpose,
we introduce a new pair of ghosts $(c,b)$ with ghost numbers $(1,-1)$
and grading numbers $(1,-1)$ and we define the new BRST operator as 
$Q' = c$. This is clearly the simplest possible choice, but it has all
the
correct properties. The space of forms has to be enlarged to
contain the ghost field $b$. In this way the vertex operators
$U^{(p,q)}$
might have even negative grading.\footnote{The situation is very
similar to the NSR superstring, where the BRST
charge $Q$ is filtered with respect to the picture number. In the Large Hilbert Space, which contains the zero mode $\xi_{0}$
of the fermion $\xi$ (obtained from the fermionization of the $\beta, \gamma$ superghosts), the BRST operator $Q$ has no cohomology. However, one can introduce a second BRST differential $Q' = \eta_{0}$
(namely, the zero mode of the conjugate momentum of $\xi_{0}$) 
with picture number $+1$ and positive ghost number, which 
is nilpotent, anticommutes with $Q$ and has no cohomology
\cite{berkovits-vafa-witten,berkovits-relating}. In addition, the
vertex operators $U^{(p,q)}$ are labeled by both the ghost number and
the picture number similarly to our superforms, which are labeled by
a ghost number and a grading number. And again similarly to our case, one can see, that by considering the sum $Q + Q'$ and restricting the functional space to finite combinations of $U^{(p,q)}$, $U^{(p)} =
\sum_{q=n_-}^{n_{+}}
U^{(p,q)}$, the cohomology is equivalent to the cohomology
of $Q$ in the Small Hilbert Space.
} It is easy to see that the cohomology of $Q'$ is empty.
                                                                                
Then, we can consider the sum of the two BRST operators
$Q + Q'$ and restrict the space of 
vertex operators $U^{(p,q)}$ to those whose grading number $n$ is 
greater or equal to the number of anticommuting ghosts that they contain, i.e. $n\geq q$. Then we have:
\begin{equation}\label{cicB}
H^{(p,q)}(Q, {\rm pure ~spinors})
\simeq H^{(p+q)}(Q+Q', n\geq q)\,.
\end{equation}
The right hand side of this equation defines correctly the spinorial cohomology discussed in \cite{swedes,CNT,CNT11d,HoweTsimpis,TsimpisDeform}. Indeed, although the cohomology of $Q$ is empty due to the constraints coming from $Q_{(n)}$ with $n\geq 1$, the introduction of $Q^{\prime}$ allows to recover the pure spinor cohomology. The technique used here has been applied to the superstring in \cite{Grassi:2001ug} in order to prove the equivalence of the BRST cohomologies with and without pure spinors \cite{Grassi:2002xv}.

We conclude the present section with a remark about the 
construction of different pictures in the pure spinor approach 
using standard methods from supergeometry. The space of forms 
$\Omega^{(p,q)}$ is not sufficient to define a sensible 
integration on supermanifolds. One has to consider the space 
of superforms of the type $\Omega^{(p,q,r)}_{\ m}$  
$$
\xi^{M_{1}} \dots \xi^{M_{p}} \l^{A_{1}} \dots \l^{A_{q+m}} 
\partial_{\lambda_{A_{1}}} \dots \partial_{\lambda_{A_{m}}} 
\prod_{k=1}^{r} \delta(C^{k}_{B} \lambda^{B})
$$
where $r$ denotes the power of delta functions in the 
generalized forms and it corresponds to the picture number. 
For a given supermanifold ${\bf R}^{(m|n)}$ 
the picture number can run from $r=0$ to $r=n+1$. 
For the two extreme values $r=0$ and $r=n+1$ one has that $n\geq 0$ and $n\leq m+1$ respectively whereas for $0<r<n+1$ the grading number $n$ can assume any value. Notice that the derivatives acting on the delta functions 
introduce negative spinorial degrees of the form. 
The differential $d$ maps in the space of forms without changing 
the picture. To change the picture, the operators $Z_{B}$ and $Y_{C}$ 
should be used. Their construction from 
pure geometrical concepts will be discussed in a forthcoming 
publication \cite{CS}.

\section{Recapitulation and discussion}

In this paper we developed the rules for calculation of one-loop amplitudes for the eleven-dimensional superparticle. Using them and the string theory counterpart of \cite{Berkovits:2004px}, we computed covariantly the following one-loop terms in the 10- and 11- dimensional effective actions: $B\wedge X_8$ in IIA/B; ${\cal F}^4$ in type I; $C\wedge X_8$ in M-theory. We also verified that the eleven-dimensional supergravity action of \cite{CJS} is reproduced by the tree-level scattering amplitudes, as should be the case for consistency. Hence, using the vertex operators $U^{(3)}$ and $U^{(1)}$ that were defined in Section~\ref{section:vop}, we can summarize the Cremmer-Julia-Scherk action by the expression~(\ref{defD}):
\begin{equation}
S_{CJS} = \langle U^{(3)} Q U^{(3)}\rangle + \langle U^{(3)}[U^{(1)}, U^{(3)}]\rangle \, , \nn
\end{equation}
where the bracket $\langle ... \rangle$ is understood with the tree-level normalization $\langle \lambda^{7}\theta^{9}\rangle=1$, recalled in Section~\ref{section:map}, and $[.\,,.]$ means taking the canonical commutators for the on-shell fields in the vertex operators.

It is possible to extend
 the present construction to multiloop amplitudes, in which case there
 will be more zero modes for the Lorentz generators. Also, as can be seen from the structure of the vertex operators in the new $b$-picture in~(\ref{newvop}), the tensors $K^{ABC}$ and $L^{ABCD}$ will contribute. As these terms are
coming with extra fermionic derivatives, the corresponding amplitude will contain higher-derivatives. It
would be interesting to try to reproduce the $D^{4}R^{4}$ terms analyzed in~\cite{GKP} from a two-loop
amplitude for the superparticle. 

However, being an approximation to the super-membrane for configurations with
constant transverse excitations, the superparticle formalism clearly cannot
grasp all information about the $\ell_{P}$-corrections to M-theory.
In particular, we saw that because the membrane vertex
operator for the anti-fields, $U^{(4)}$, 
 does not enter in the definition of the tree and loop
amplitudes for the superparticle, it is not possible to deduce any information
about the $(\ell_{P})^3$-deformations of the four-form field strength
addressed in \cite{TsimpisDeform,Witten:1996md}.
 
Extension of the present work to the full supermembrane is not obvious. First of all, the BRST operator for the M$2$-brane requires new secondary constraints \cite{membr}:
\begin{equation}
(\lambda \Gamma^{MN}\lambda)\, \Pi_{JM}(g,C_{3})=0\,,\quad \lambda_{A}\partial_{J}\lambda^{A}=0 \ \ {\rm with} \quad
 \Pi_J{}^{M}=E_{\Omega}{}^{M}\partial_{J} Z^{\Omega} \ \ {\rm and} \quad J=1,2 \,,
\end{equation}
 involving the transverse excitations of the membrane. One can view the first constraint as a restriction not on the pure spinor but instead on the background metric $g$ and three-form $C_{3}$ in which the membrane evolves. It is interesting to notice that the last equation is not only algebraic but involves the dynamics of the pure spinors. 

Another expected difficulty stems from the complicated nonlinear dynamics of the membrane excitations. In particular, even BPS amplitudes are not expected to be easy to understand based on the work of \cite{Sugino:2001iq} which showed that, although the counting of membrane instantons wrapping $T^{3}$ can be obtained by computing the partition function of an  associated 3d Matrix model, the model does not localize on the semi-classical configurations of the long membrane wrapping the $T^{3}$ torus.


\section*{Acknowledgments}
The authors would like to thank the organizers of the second Simons Workshop in Stony Brook, 2004, for 
providing a very nice environment, where this paper could be completed, and for financial support.
We would like to thank Nathan Berkovits for many useful discussions and for
reading the manuscript. We are also grateful to M. Ro\v{c}ek, W. Siegel and P. van Nieuwenhuizen for
positive feedback. P.A.G. thanks the institute IHES, Bures-sur-Yvette,
for hospitality during the initial stages of this work and also L. Castellani
and A. Lerda for useful conversations. P.V. thanks the LPTHE of Jussieu and
the Physics Department of Neuch\^{a}tel University for hospitality.

The research of L.A. was supported in part by DOE grant DE-FG02-95ER40899. The
research of P.A.G. was partially supported by NSF-grant PHY-0354776. P.V. was
partially supported by the EU networks HPRN-CT-2000-00148, and
HPRN-CT-2000-00131.


\appendix\section{Fierz identities and Gamma matrix manipulations in 11d} \label{section:Fierz}

In this appendix we write down several Fierz identities that we need in the main text. We refer to \cite{japanFierz} for an exhaustive list of the Fierz identities in eleven dimensions. 

As Berkovits' pure spinors satisfy only (\ref{puredef}) but not $\l^A\Gamma^{MN}_{AB}\l^B = 0$, the Fierz identity between two $\lambda$'s takes the form: 
$$ 
32\, \lambda^{(A} \lambda^{B)} = \frac{1}{2!} \, \left(\Gamma^{MN}\right)^{AB} (\lambda \Gamma_{MN}\lambda) + \frac{1}{5!} \left(\Gamma^{M_1\cdots M_5}\right)^{AB} (\lambda \Gamma_{M_1\cdots M_5}\lambda) \, .
$$ 

The $d^{23}N$-measure (\ref{dNmeasure}) is expanded on products of bilinears of $\lambda^A$ made from $\Gamma$-matrices with two, five and six indices. The relevant Fierz identities are:  
\begin{eqnarray}\label{Mtwo} 
\left(\Gamma^M\right)_{(AB} \left(\Gamma_{MN}\right)_{CD)}&=&0 \, ,\\ 
\label{Mfive}\left(\Gamma_M\right)_{(AB} \left(\Gamma^{MN_1\cdots N_4}\right)_{CD)}&=&6\left(\Gamma^{[N_1N_2} \right)_{(AB} \left(\Gamma^{N_3N_4]} 
\right)_{CD)} \, ,\\ 
 \left(\Gamma_M\right)_{(AB} \left(\Gamma^{MN_1\cdots N_5}\right)_{CD)}&=&-\frac{1}{2}\delta_{(AB} \left(\Gamma^{N_1\cdots N_5}\right)_{CD)} -\frac{5}{2} \left(\Gamma^{[N_1}\right)_{(AB} \left(\Gamma^{N_2\cdots N_5]}\right)_{CD)} \nn\\  
 &+& 5\left(\Gamma^{[N_1N_2}\right)_{(AB} \left(\Gamma^{N_3\cdots N_6]}\right)_{CD)} \, ,\\ 
\label{GammaSix} 
\left(\Gamma^{M_1M_2}\right)_{(AB} \left(\Gamma^{N_1\cdots N_6}\right)_{CD)}&=&-\frac{1}{4}\delta_{(AB} \left(\Gamma^{M_1M_2N_1\cdots N_6}\right)_{CD)} \nn\\ 
& +&\frac{5}{2} \left(\Gamma^{([M_1M_2N_1N_2}\right)_{(AB} \left(\Gamma^{N_3\cdots N_6]}\right)_{CD)} \, .
\end{eqnarray} 
(\ref{Mtwo}) is known as the M$2$-brane identity and (\ref{Mfive}) $-$ as the M$5$-brane identity. These Fierz identities immediatly imply that: 
\begin{eqnarray}\label{Idone} 
\left(\lambda \Gamma_M\right)_A \left(\l \G^{MN}\l\right)&=&0 \, ,\\ 
\label{Idtwo}\left(\lambda\Gamma_M\right)_A \left(\lambda\Gamma^{MN_1\cdots N_4}
\lambda\right)&=&\left(\lambda\Gamma^{[N_1N_2} \right)_A \left(\lambda\Gamma^{N_3N_4]}\l\right) \, ,\\ 
\label{Idthree}(\lambda \Gamma^{M_1M_2}\lambda) (\lambda \Gamma^{N_1\cdots N_6}\lambda) &=&0 \, ,\\
\label{Idfour} 
\lambda_A \left(\lambda\Gamma^{M_1\cdots M_6}\lambda\right)&=&-6\left(\lambda\Gamma^{[M_1}\right)_A \left(\lambda\Gamma^{M_2\cdots M_6]}\lambda\right) \nn\\
&&+15\left(\lambda\Gamma^{[M_1M_2}\lambda\right) \left(\lambda\Gamma^{M_3\cdots M_6]}\right)_A \, .
\end{eqnarray} 
From (\ref{Idthree}) it follows that the measure factor for $dN$ in (\ref{dNmeasure}) cannot contain any terms of the form $(\l \Gamma_{[2]}\l)\, (\l \Gamma_{[6]}\l)$, where as in the main text we denote by $\Gamma_{[n]}$ the antisymmetric product of $n$ Gamma-matrices.

It is also useful to recall the explicit form of the $t_8$-tensor:
\bea \label{teightI}
t_8^{m_1 m_2 n_1 n_2 p_1 p_2 q_1 q_2} &=& -2 (\d^{m_1 n_2} \d^{m_2 n_1} \d^{p_1 q_2} \d^{p_2 q_1} + \d^{n_1 p_2} \d^{n_2 p_1} \d^{m_1 q_2} \d^{m_2 q_1} + \d^{m_1 p_2} \d^{m_2 p_1} \d^{n_1 q_2} \d^{n_2 q_1}) \nn \\ 
&+& 8 (\d^{m_1 q_2} \d^{m_2 n_1} \d^{n_2 p_1} \d^{p_2 q_1} + \d^{m_1 q_2} \d^{m_2 p_1} \d^{p_2 n_1} \d^{n_2 q_1} + \d^{m_1 n_2} \d^{m_2 p_1} \d^{p_2 q_1} \d^{q_2 n_1}) \nn \\ 
&+& \text{anti-sym. in every index pair} \, . 
\eea 
In $d=10$ and $d=11$ it can be represented in terms of gamma-matrices as follows: 
\begin{eqnarray}\label{teightII}
2\Delta\, t_8^{m_1 m_2 n_1 n_2 p_1 p_2 q_1 q_2} &=&
\text{tr}\left(\Gamma^{m_{1}m_{2}}\Gamma^{m_{3}m_{4}}\Gamma^{m_{5}m_{6}}\Gamma^{m_{7}m_{8}}\right)
+\text{tr}\left(\Gamma^{m_{5}m_{6}}\Gamma^{m_{1}m_{2}}\Gamma^{m_{3}m_{4}}\Gamma^{m_{7}m_{8}}\right)\\
\nonumber &+&\text{tr}\left(\Gamma^{m_{3}m_{4}}\Gamma^{m_{5}m_{6}}\Gamma^{m_{1}m_{2}}\Gamma^{m_{7}m_{8}}\right)
-\frac{2}{\Delta}
\text{tr}\left(\Gamma^{m_{1}m_{2}}\Gamma^{m_{3}m_{4}}\right) 
\text{tr}\left(\Gamma^{m_{5}m_{6}}\Gamma^{m_{7}m_{8}}\right)\\
\nonumber &-&\frac{2}{\Delta}
\text{tr}\left(\Gamma^{m_{1}m_{2}}\Gamma^{m_{5}m_{6}}\right) 
\text{tr}\left(\Gamma^{m_{3}m_{4}}\Gamma^{m_{7}m_{8}}\right)
-\frac{2}{\Delta}
\text{tr}\left(\Gamma^{m_{1}m_{2}}\Gamma^{m_{7}m_{8}}\right) 
\text{tr}\left(\Gamma^{m_{3}m_{4}}\Gamma^{m_{5}m_{6}}\right)\, ,
\end{eqnarray}
where $\Delta$ is the dimension of the Clifford algebra. 

\section{SO(8) parametrization of the 11d Pure Spinors}\label{SOeight}

In this appendix we explain the $SO(8)$ decomposition of an eleven-dimensional pure spinor, which makes manifest the number of independent components it has.

In Section~\ref{section:map} we introduced a commuting spinor $\l^{\prime} = (\l'_1 , \l'_2)$ satisfying the constraint $\l^{\prime} \Gamma^+ \l^{\prime} = 0$, i.e. $\lambda'_1 \gamma^+ \lambda'_1 + \lambda'_2 \gamma^+ \lambda'_2 = 0$. This constraint can be solved in the following way: decompose $\lambda'_1$ into $s_a$ and $s_{\dot a}$ \,,  
belonging to the $SO(8)$ representations $8_c$ and $8_s$ respectively. Then the equation \, $\lambda'_1 \gamma^+ \lambda'_1 + \lambda'_2 \gamma^+ \lambda'_2 = 0$ \, becomes $s_a s_a = v^+$, where $v^+ \equiv \lambda'_2 \gamma^+ \lambda'_2$ (the $SO(8)$ indices are raised and lowered with the metrics $\delta_{ab}$ and $\delta_{\dot a \dot b}$).   
The equation is preserved by $SO(7)$ rotations:  
using the triality to view $s_a$ as a vector of $SO(8)$ and decomposing it into the coset representation $S^{7} \times SO(7)$, one can fix its  
8$^{th}$ component to be equal to $\sqrt{ v^+ - s_i s_i}$ \,,   
where $s_i$ is the $SO(7)$-part of $v$. Then, using the similarity transformation of Section \ref{section:map}, one can find the BRST operator  
$\lambda^\alpha_1 d_{1\alpha} + \lambda_{2\alpha} d^\alpha_{2}$ \,, where the 11d pure spinor is given by  
\begin{eqnarray} 
\lambda = (\lambda_1, \lambda_2) = \Big( (\sqrt{v^+ - s_i^2}, s_i), (0, 0), \lambda'_2 \Big). 
\end{eqnarray} 
The first two entries are the $s_a$ of $\lambda'_1$, and the third and the fourth ones, (0,0),  
are the 1 + 7 components of $s_{\dot a}$  with respect to $SO(7)$ using again triality.  
Finally, $\lambda'_2$ is completely free. The counting of degrees of freedom gives 7 + 16 = 23 (complex),   
which is exaclty the number of independent components of a pure spinor in 11d.

Notice that the above computation is parallel to the 10d one of Berkovits. Indeed, in ten dimensions the $SO(9,1)$ spinor $\lambda_\alpha$ is reduced to two $SO(8)$ spinors,  
$\tilde{s}_a$ and $\tilde{s}_{\dot a}$. Then one imposes the pure spinor constraint and  finds that $\tilde{s}_a \tilde{s}_a = 0$. So the spinor $\gamma^+ \lambda = \tilde{s}_a$ is null, i.e. $\lambda \gamma^+ \lambda = 0$. Then  
one can decompose the rest under a $SU(4)$ subgroup of $SO(8)$, which  
preserves the null property of $\tilde{s}_a$. Therefore the resulting solution for a pure spinor is  
$$\lambda = \{ \tilde{s}_a, s_A, 0 \} \, ,$$ 
where $s_A$ is a generic spinor of $SU(4)$. Hence there are 4 + 7 = 11 independent components, coming respectively from $s_A$ and $\tilde{s}_a$. Comparing with the above 11d case, we see that $s_A$ replaces $\lambda'_2$ whereas $\tilde{s}_a$ replaces $(s_a, 0)$ with $s_a$ $-$ a spinor of $S^7$. 


\end{document}